\documentclass[a4paper,11pt]{article}
\pdfoutput=1
\usepackage[utf8]{inputenc}
\usepackage{latexsym}
\usepackage{epstopdf}
\usepackage{epsfig}
\usepackage{graphicx}
\usepackage{amsmath}
\usepackage{multirow}
\usepackage{subfigure}
\usepackage{a4wide}
\usepackage{cite}
\newcommand{\be}{\begin{equation}}
\newcommand{\ee}{\end{equation}}
\newcommand{\ba}{\begin{eqnarray}}
\newcommand{\ea}{\end{eqnarray}}
\newcommand{\bac}{\begin{array}{c}}
\newcommand{\eaa}{\end{array}}
\newcommand{\baz}{\begin{array}{cc}}
\newcommand{\mathsym}[1]{{}}

\newcommand{\bad}{\begin{array}{ccc}}

\newcommand{\bi}{\begin{itemize}}
\newcommand{\ei}{\end{itemize}}
\newcommand{\bmt}{\begin{pmatrix}}
\newcommand{\emt}{\end{pmatrix}}
\newcommand{\bt}{\begin{tabular}}
\newcommand{\et}{\end{tabular}}

\newcommand{\benu}{\begin{enumerate}}
\newcommand{\eenu}{\end{enumerate}}

\newcommand{\bav}{\begin{array}{cccc}}

\begin{document}

\title {\Large\bf{ Neutrino Mass, Coupling Unification, Verifiable Proton Decay, Vacuum
      Stability and WIMP Dark Matter in SU(5)}}

\author{\bf Biswonath Sahoo$^{*}$, Mainak Chakraborty$^{**}$,  M.K. Parida $^{\dagger}$   \\
 Centre of Excellence in Theoretical and Mathematical Sciences, \\
Siksha \textquoteleft O\textquoteright Anusandhan (Deemed to be University),  \\
Khandagiri Square, Bhubaneswar 751030, Odisha,  India}

\maketitle

\begin{abstract}
Nonsupersymmetric minimal SU(5) with Higgs representations ${24}_H$ and
$5_H$ and standard fermions in ${\bar 5}_F\oplus {10}_F$ is well known 
for its failure in unification of gauge couplings and lack of predicting 
neutrino masses. Like standard model, it is also affected by the instability 
of the Higgs scalar potential. We note that extending the Higgs sector by 
${75}_H$ and ${15}_H$  not only leads to the popular type-II seesaw ansatz
for neutrino masses with a lower bound on the triplet mass 
$M_{\Delta} > 2\times 10^9$ GeV, but also achieves precision unification
of gauge couplings without proliferation of non-standard light Higgs
scalars or fermions near the TeV scale. Consistent with recent LUX-2016
lower bound, the model easily accommodates a singlet scalar WIMP dark matter 
near the TeV scale which resolves the  vacuum  stability issue even after 
inclusion of heavy triplet  threshold effect. We estimate proton lifetime 
predictions for $p\to e^+\pi^0$ including uncertainties due to input parameters 
and threshold effects due to superheavy Higgs scalars and superheavy  
$X^{\pm 4/3},Y^{\pm 1/3}$ gauge bosons. The predicted lifetime is noted to be
verifiable at Super Kamiokande and Hyper Kamiokande experiments.         
\end{abstract}
\noindent{${}^*$email:sahoobiswonath@gmail.com}\\
\noindent{${}^{**}$email:mainak.chakraborty2@gmail.com}\\
\noindent{${}^{\dagger}$email:minaparida@soa.ac.in}

\section{Introduction}\label{sec:intr}
Standard model (SM) of strong and electroweak interactions has been established by numerous experimental tests, yet evidences on neutrino mass
 \cite{Salas:2017,schwetz,forero,fogli,gonzalez}, the phenomena of dark matter
 \cite{zwicky,spergel:2007,einasto,blumenthal,angle,strigari,Rubin:1970,Clowe:2006,XENON,LUX:2013,LUX:2016,Fermi-LAT,IceCube:2013,IceCube:2014,IceCube:2016,IceCube-ANTA:2015,LHC:2017,Sirunyan:2017}, and 
 baryon asymmetry of the universe (BAU) \cite{spergel:2007,spergel:2003,komatsu,hindshaw,Planck15}
call for beyond standard model (BSM) physics. It is well known that
grand unified theories (GUTs)
\cite{JCP:1974,Georgi-Glashow:1974,Georgi:1974,Fritzsch-Minkowski:1975,Langacker:1981,Slansky:1979,Nath-Perez:2007}
are capable of addressing a number of  limitations of
the SM  effectively. There are interesting theories on neutrino mass
generation mechanisms \cite{nurev1,nurev2} based upon various seesaw
mechanisms such as type-I, type-II,
type-III\cite{type-I,Valle:1980,type-II,Ma-Us:1998,type-III,Ma:1998,Bajc-gs:2007},
linear \cite{Linear} and inverse \cite{Inverse,Valle-LFV,Inverse1}. Interesting models for Dirac neutrino
mass origin of the neutrino oscillation data have been also proposed
\cite{Valle:Dirac}. In the absence of experimental evidence of
supersymmetry so far, non-supersymmetric (non-SUSY) GUTs are being
extensively exploited by reconciling to the underlying gauge
hierarchy problem through finetuning
\cite{Weinberg:2007,Barr:2010}. Higher rank GUTs like SO(10)  and
$E_6$ can not define a unique symmetry breaking path to the SM gauge
theory because of large number of possibilities with one and more
intermediate symmetry breakings consistent with electroweak precision data
on $\sin^2\theta_W(M_Z), \alpha_S(M_Z)$, and $\alpha (M_Z)$ \cite{cmp:1984}. On the
other hand, the rank-4 minimal 
SU(5)\cite{Georgi-Glashow:1974} with Higgs representations ${5}_H$ and
${24}_H$ defines only one unique symmetry breaking path to the
standard model
\begin{equation}
SU(5)\to SM .    \label{eq:su5br}
\end{equation}
Type-I seesaw \cite{type-I,Valle:1980} needs non-standard heavy right-handed neutrino,  linear and
inverse
seesaw   both need nonstandard fermions and scalars, and type-III seesaw
\cite{nurev1,nurev2,type-III,Ma:1998,Bajc-gs:2007} needs only non-standard
fermionic extension for their implementation. Out of these popular
seesaw mechanisms, Type-II seesaw mechanism is the one which 
needs a heavy non-standard triplet scalar 
\cite{Valle:1980,type-II,Ma:1998,Ma-Us:1998}. With a second triplet
scalar, it is also capable of predicting baryon asymmetry of the universe
\cite{Ma-Us:1998} which is one of the main motivations behind
this investigation.  This neutrino mass generation mechanism, gauge coupling
unification, dark matter, and vacuum stability  are
the focus of the present work.
    
Like the minimal SM, with its $15$ fermions per generation and the
standard Higgs doublet $\phi(2,1/2,1)$, the minimal SU(5) with Higgs representations ${5}_H$ and
${24}_H$ predicts neutrinos to be massless subject to a tiny  ${\cal
  O} (10^{-5})$ eV contribution due to nonrenormalizable Planck-scale
effect which is nearly $4$ orders smaller than the requirement of
neutrino oscillation data. As the particle spectrum below the GUT
symmetry breaking scale is identically equal to the SM spectrum, like
SM, the minimal GUT fails to unify gauge couplings \cite{Langacker:1993,Ma:2009,Stella:2016}. 
Also it predicts instability of the  Higgs quartic coupling at mass scales
$\mu \ge 5\times 10^{9}$ GeV \cite{Elias-Miro:2012,Lebedev:2013,Moroi:2018} after which
the coupling continues to be increasingly negative at least up to the unification scale. 

 A number of interesting models have
been suggested for coupling unification by populating the grand desert
and  for enhancing proton lifetime predictions
\cite{Bajc-gs:2007,Ma:2005,Dorsner}. In  these models a number of fermion or
scalar masses below the GUT scale have been utilised to achieve unification. Interesting
possibility of type-III seesaw \cite{Bajc-gs:2007} with experimentally verifiable dilepton
production  \cite{Keung-gs:1982} at LHC has been also investigated.
 
The other shortcoming  of minimal non-SUSY SU(5)  is its inability to predict
  dark matter     
 which appears to belong to two distinct categories: (i) The 
 weakly interacting massive particle (WIMP) dark matter of  bounded mass $ < 100 $ TeV,  (ii) The decaying dark matter which
 has been suggested to be a possible source of PeV energy IceCube neutrinos.  

In this work we implement a novel  mechanism for coupling unification and 
neutrino masses  together. When SU(5) is extended by the addition of its Higgs representations ${75}_H$ and ${15}_H$, it achieves two objectives: (i) Neutrino mass and mixing generation through type-II seesaw mechanism, and (ii) Precision gauge coupling unification with experimentally accessible proton lifetime.

But this does not cure the vacuum instability problem persisting in
the model as well as  the need for   
 WIMP dark matter prediction. Out of these two, as we note in this work, when the dark matter prediction is successfully inducted into the model, the other problem on vacuum stability is automatically resolved. 

 In contrast to the popular belief on low proton lifetime prediction
 of the minimal SU(5) \cite{Langacker:1981},  we estimate new precise
 and enhanced predictions of this model including
threshold effects
\cite{Weinberg:1980,Hall:1981,Ovrut:1982,mkp:1987,mkp-cch:1989} of
heavy particles near the GUT scale. Predicted lifetimes are found
to be within the  accessible ranges of Superkamiokande
and Hyperkamiokande experimental search programmes \cite{Abe:2017}.\\

This paper is organised in the following manner. In Sec.\ref{sec:numass}
we discuss neutrino mass generation mechanism in extended SU(5). 
Sec.\ref{sec:unif} deals with the problem of gauge coupling unification. In
Sec.\ref{sec:plife} we make proton lifetime prediction including
possible uncertainties. Embedding WIMP scalar DM in SU(5) is discussed
in Sec. \ref{sec:wimpdm} with a brief outline on the current experimental
status.
Resolution of  vacuum stability issue is explained in
Sec.\ref{sec:vacstab}. We summarise and conclude in
Sec. \ref{sec:sum}.  Renormalisation group equations for gauge and
Higgs quartic couplings are discussed in the Appendix.

\section{ Neutrino Mass Through Type-II Seesaw in SU(5)
} \label{sec:numass}
As noted in Sec.\ref{sec:intr}, in contrast to many possible
alternative symmetry breaking paths to SM from non-SUSY SO(10) and
$E_6$ \cite{cmp:1984} , SU(5) predicts only one symmetry breaking path
which enhances its verifiable predictive capability. 
Fifteen SM fermions are placed in two different SU(5) representations
\begin{eqnarray}
&&{\overline 5}_F={\begin{pmatrix}d^C_1\\
d^C_2\\
d^C_3\\
e^-\\
-\nu_e \end{pmatrix}}_L, \nonumber\\
&&{10}_F={\begin{pmatrix} 0 & u_2^C & -u_3^C & u_1 & d_1\\
-u_2^C& 0 & u_1^C & u_2 & d_2\\
u_3^C &-u_1^C& 0 & u_3 & d_3\\
-u_1 & -u_2 & -u_3 & 0 & e^C\\
-d_1 & -d_2 & -d_3 & -e^C & 0 \end{pmatrix}}_L. \label{eq:fermi5}
\end{eqnarray}
Lack of RH$\nu$ in these representations  gives vanishing Dirac neutrino mass and
vanishing Majorana neutrino mass at renormalizable level. Planck-scale
induced small Majorana masses can be generated through
non-renormalizable ${\rm dim.}5$ interaction

\begin{equation}
-{\cal L}_{NR}= \frac{\kappa_{ij}}{M_{\rm Planck}}{\bar 5}_{F_i}{\bar 5}_{F_j}{5}_H{5}_H +
h.c. \label{eq:NRmnu}
\end{equation}

 leading to $m_{\nu} \sim 10^{-5}$ eV which is too low to explain
neutrino oscillation data.
 Mechanism of Dirac neutrino mass generation
has been discussed \cite{Valle:Dirac} matching the neutrino oscillation data.
Using extensions of the minimal GUT type-III seesaw origin of neutrino mass has been discussed  where the nonstandard fermionic triplet $\Sigma_F(3,0,1)$
mediates the seesaw. This model can be experimentally tested by the
production of like-sign dilepton signals at LHC. 

Type-II seesaw mechanism for neutrino mass \cite{type-II,Ma-Us:1998} does not need any non-standard fermion, but needs only the non-standard left-handed Higgs scalar triplet $\Delta_L(3,-1,1)$  with $Y=-2$ which directly couples with the a dilepton pair. It also directly couples to standard  Higgs doublet $\phi$. As such the standard Higgs VEV can be transmitted as a small induced VEV generating Majorana mass term for the light neutrinos.
As this $\Delta_L(3,-1,1)$ is contained in the symmetric SU(5) scalar representation ${15}_H$, the scalar sector of the minimal GUT needs to include ${15}_H$ in addition to ${5}_H$ and ${24}_H$.

\begin{figure}[h!]
\begin{center}
\includegraphics[width=6cm,height=5cm,angle=0]{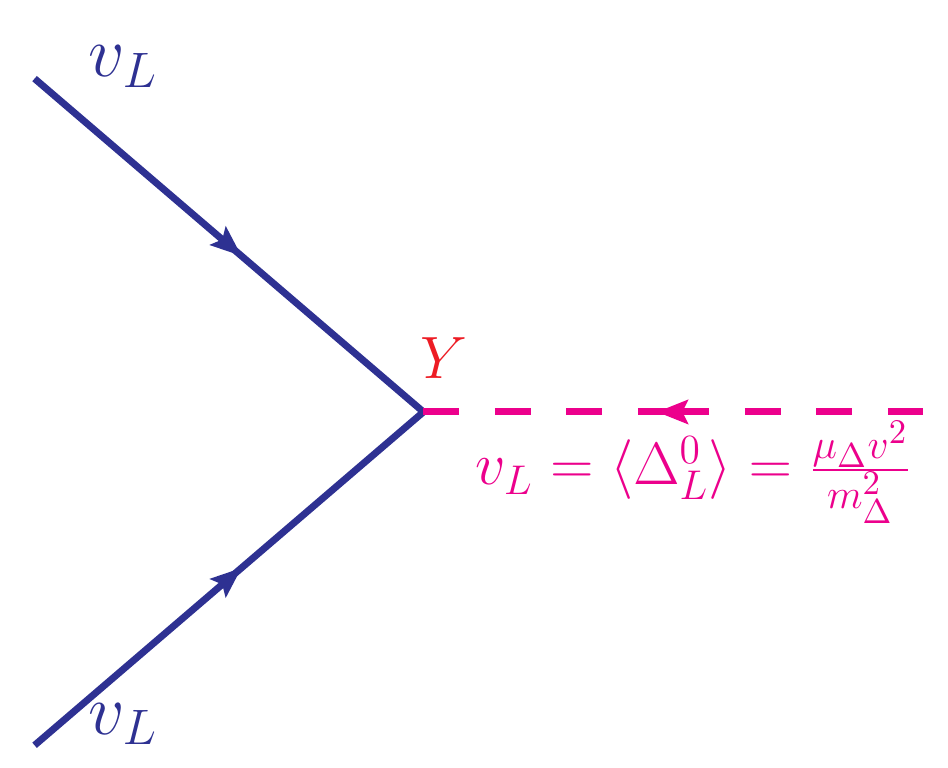}
\hspace{1.0cm}
\includegraphics[width=6cm,height=5cm,angle=0]{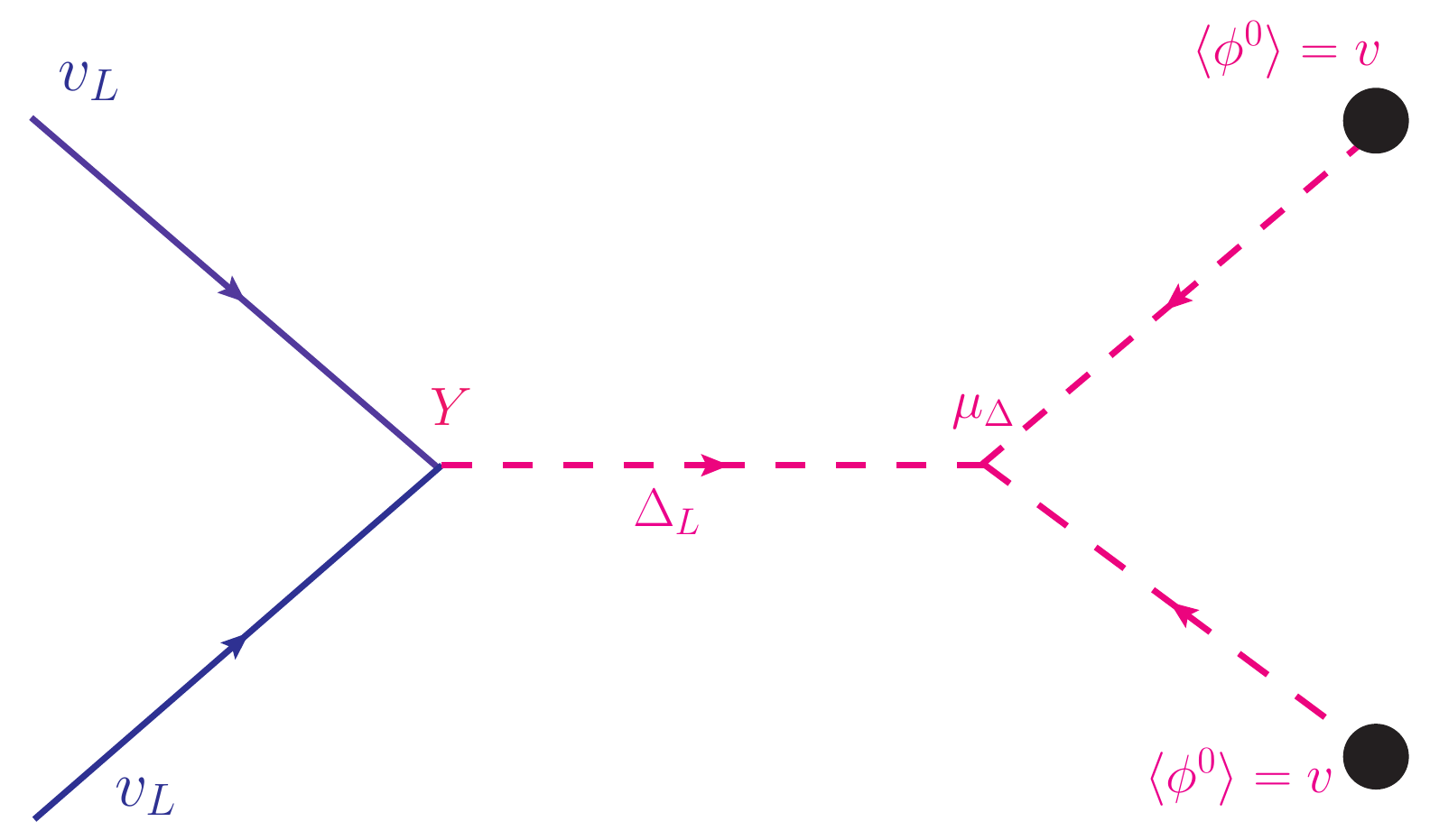}
\caption{Schematic representation of generation of Type-II term
  corresponding to  eq.(\ref{eq:Yukll}) (left panel) and combination of 
eq.(\ref{l2})(right panel) where dashed line as triplet propagator supplies the damping factor $M_{\Delta}^{-2}$ to the induced vev $v_L$. }
\label{feyn1}
\end{center}
\end{figure}

The Yukawa Lagrangian
\begin{equation}
-\mathcal{L}^{(II)}=l_{L_i}^T C i \tau_2 Y_{ij}  {(\frac{\vec{\tau}.\vec{\Delta_L}}{\sqrt{2}})}^{\dagger} l_{L_j}+ h.c. \label{eq:Yukll}
\end{equation}
combined with the relevant part of the Higgs potential
\begin{equation}
{\cal V}_{II}= M_{\Delta}^2 Tr [\Delta_L^\dagger \Delta_L]+
\mu_\Delta \tilde{\phi}^\dagger(\frac{\vec{\tau}.\vec{\Delta_L}}{\sqrt{2}})
\phi +h.c, \label{l2}
\end{equation}
gives rise to the type-II seesaw contribution. In our notation 
$l_{L_i}^T=(\nu_{L_i},~e_{L_i})$~($i=$  generation index), $\phi^T=(\phi^+,\phi^0)$ which are the lepton and scalar doublet of $SU(2)_L$.
Here $\tilde{\phi}=i\tau_2\phi^\ast$, $\vec{\tau}=(\tau_1,\tau_2,\tau_3)$ ($\tau_i$ are the $2\times2$ Pauli spin matrices) and, similarly, the scalar triplet
$\Delta_L$ in the adjoint representation of $SU(2)_L$ is expressed as
$\vec{\Delta_L}=(\Delta_L^1,\Delta_L^2,\Delta_L^3)$. The Majorana type Yukawa
coupling $Y$ is a  $3\times3$ matrix in flavor space and $C$ is the charge conjugation matrix. Then
\begin{eqnarray}
(\frac{\vec{\tau}.\vec{\Delta_L}}{\sqrt{2}})&=&\frac{1}{\sqrt{2}}(\tau_1 \Delta_L^1 +  \tau_2 \Delta_L^2 + \tau_3 \Delta_L^3)\nonumber\\
                                            &=&\left( \begin{array}{cc}
     \frac{\Delta^+}{\sqrt{2}} & \Delta^{++} \\ \Delta^0 & -\frac{\Delta^+}{\sqrt{2}}  \\ 
    \end{array}\right)_L
\end{eqnarray}                              
where different components are given by
\begin{equation}
\Delta_L^0=\frac{1}{\sqrt{2}}(\Delta_L^1+i\Delta_L^2),~~\Delta_L^+=\Delta_L^3,~~\Delta_L^{++}= \frac{1}{\sqrt{2}}(\Delta_L^1-i\Delta_L^2)
\end{equation}
A diagrammatic representation for type-II seesaw generation of neutrino mass is shown in Fig.\ref{feyn1}. 
From the Feynman diagram shown in this figure \ref{feyn1} the induced VEV of the scalar triplet is
\be
v_L=\frac{\mu_{\Delta} v^2}{2 M_\Delta^2}.\label{eq:vl}
\ee
leading to the type-II seesaw formula
\begin{equation}
m_\nu =2Yv_L \label{t2} .
\end{equation}
It is necessary to explain the origin of the $B-L$ breaking scale
$\mu_{\Delta}$ that occurs in eq.(\ref{eq:Yukll}), eq.(\ref{l2}), and
eq.(\ref{eq:vl}) as well as the Feynman diagram of Fig. \ref{feyn1}.
SU(5) invariance permits the triplet coupling
$\mu_{\Delta}{15}_H^{\dagger}{5}_H{5}_H$ leading to SM invariant
coupling $\mu_{\Delta}\Delta_L\phi\phi$. Therefore, in one approach,  $\mu_{\Delta}$ may
be treated as explicitly lepton number violating parameter. Alternatively, it is also possible to attribute a spontaneous lepton number violating origin to this parameter.   
Since the SM model gauge theory has to remain unbroken down to the
electroweak scale, the lepton numder violating scale can
be generated by the VEV of a  Higgs scalar that transforms as a
singlet under SM. Such a singlet $S_{BL}(1,0,1)$ carrying
$B-L=-2$  occurs in the Higgs representation
${50}_H$ \cite{Slansky:1979,pnsa:2016}. The part of the SU(5) invariant potential that generates this scale is
\begin{equation}
V_{BL}=\lambda_5{50}_H{15}_H{5}_H{5}_H+
h.c. \label{eq:vbl}
\end{equation}
leading to $\mu_{\Delta}=\lambda_5<S_{BL}>$.
The $U(1)_{B-L}$ symmetric origin of $\mu_{\Delta_L}$ becomes more transparent if one treats SU(5) as the remnant of $SU(5)\times U(1)_{B-L}$ or higher rank GUTs like SO(10) and $E_6$. 
If unification constraint as discussed below is ignored,
the order of magnitude of  $\mu_{\Delta}$  can be anywhere between $\cal O (M_{\rm W}) - \cal O
(M_{\rm Planck})$. But as we will find in the subsequent sections, gauge coupling
unification in the present SU(5) framework imposes the lower bound
$\mu_{\Delta}\simeq M_{\Delta} \ge 10^{9.23}$ GeV .  
\subsection{ Type-II Seesaw Fit to the Neutrino Oscillation Data}
\subsubsection{ Neutrino Mass Matrix from Oscillation Data}
The effective light neutrino mass matrix $(m_\nu)$ is diagonalised by a unitary matrix $U$ (in PMNS parametrisation which is written as $U_{\rm PMNS}$)
and yields three mass eigenvalues $(m_1,m_2,m_3)$. The light neutrino mass matrix ($m_\nu$) can be reconstructed as 
\begin{equation}
m_\nu = U_{PMNS}~diag(m_1, m_2, m_3) U_{PMNS}^T ,\label{mnu}
\end{equation}
where  PMNS matrix is parameterised using the  PDG convention\cite{PDG:2012} as
\begin{equation}
 U_{\rm{PMNS}}= \left( \begin{array}{ccc} c_{12} c_{13}&
                      s_{12} c_{13}&
                      s_{13} e^{-i\delta}\cr
-s_{12} c_{23}-c_{12} s_{23} s_{13} e^{i\delta}& c_{12} c_{23}-
s_{12} s_{23} s_{13} e^{i\delta}&
s_{23} c_{13}\cr
s_{12} s_{23} -c_{12} c_{23} s_{13} e^{i\delta}&
-c_{12} s_{23} -s_{12} c_{23} s_{13} e^{i\delta}&
c_{23} c_{13}\cr
\end{array}\right) 
diag(e^{\frac{i \alpha_M}{2}},e^{\frac{i \beta_M}{2}},1)
\end{equation}
where $s_{ij}=\sin \theta_{ij}, c_{ij}=\cos \theta_{ij}$ with
$(i,j=1,2,3)$, $\delta$ is the Dirac CP phase and $(\alpha_M,\beta_M)$ are Majorana phases.
\paragraph{}
Here we present our numerical analysis within $3\sigma$ and $1\sigma$
limits of experimental data. As we do not have any experimental information 
about Majorana phases, they are varied in the whole $2\pi$ interval randomly. From the set of randomly generated values we pick only one 
set of $(\alpha_M,\beta_M)$ and use them for our numerical
estimations. The procedure adopted here can be repeated to derive
corresponding  solutions for the Majarana coupling matrix $Y$ for other sets
of randomly chosen Majorana phases.
Although very recently  $3\sigma$  and $1\sigma$ 
limits of Dirac CP phase has been announced \cite{Salas:2017},
we prefer to use only their central value as an example. For our present analysis
we choose a single set of $(\alpha_M,\beta_M)$ from a number of sets
derived by random sampling and also a single value of $\delta$ close
to the best fit value.  For our anlysis
all possible values of the solar and atmospheric mass squared
differences and mixing angles have been taken
which lie within the $3\sigma$ (or $1\sigma$) limit of the oscillation data as determined by recent global analysis\cite{Salas:2017}. Summary of the global
analysis is presented in the Table \ref{osc} below. At first we analyze the limits imposed on the neutrino Yukawa couplings by $3\sigma$ 
oscillation constraints taking into account both the mass ordering of
light neutrinos, normal ordering (NO) and inverted ordering (IO). In this case we use only one fixed value of the lightest
neutrino mass eigenvalue and the other two mass eigenvalues are calculated using the experimental values of the mass squared differences.
In this $3\sigma$ case we represent the bounds on the elements of Yukawa matrix in a tabular form. Later we proceed to estimate the 
bounds on the $Y$ matrix elements imposed by $1\sigma$ experimental constraints of oscillation observables. In this analysis instead of fixed 
lightest neutrino mass eigenvalue, we vary it in the range $(0-0.2)$ eV. The other two mass eigenvalues are calculated using $1\sigma$
ranges of solar and atmospheric mass squared differences. As already
explained we use a single set of randomly chosen $(\alpha_M,\beta_M)$
and the central value of $\delta$  
 quoted in the Table \ref{osc}. The variation of $Y$ matrix elements
 is expressed in terms of their moduli ( $|Y_{ij}|$) and the
 corresponding phases $(\phi_{ij})$) 
with $m_1$ is shown graphically in Figs. \ref{y_mod1}, \ref{y_mod2},
\ref{phase1} in the NO case. It is clear from the plots that for each single value of $m_1$ there is a band 
of allowed values of $|Y_{ij}|$  and $\phi_{ij}$. This band signifies the $1\sigma$ allowed range of the corresponding matrix element for that single 
value of $m_1$. To represent the $1\sigma$ bounds in a more
transparent manner we produce another set of plots  as in
Fig. \ref{y_mod11} and Fig. \ref{phase22} where we show 
the allowed values of $|Y_{ij}|$ and $\phi_{ij}$ for a fixed value of $m_1$. It is to be noted that in this present work graphical representation
is done for normally ordered light neutrinos only. Similar kind of exercise can be carried out for inverted mass ordering also.

\begin{table}[!h]
\caption{Input data from neutrino oscillation experiments \label{osc} 
\cite{Salas:2017}}
\label{input}
\begin{center}
\begin{tabular}{|c|c|c|c|c|}
\hline
{ Quantity} & {best fit values} &{ $3\sigma$ ranges}&{ $2\sigma$ ranges} & { $1\sigma$ ranges}\\
\hline
$\Delta m_{21}^2~[10^{-5}eV^2]$ & $7.55$ & $7.05-8.14$ & $7.20-7.94$& $7.39-7.55$\\
$|\Delta m_{31}^2|~[10^{-3}eV^2](NO)$ & $2.50$ & $2.41-2.60$ & $2.44-2.57$& $2.47-2.53$\\
$|\Delta m_{31}^2|~[10^{-3}eV^2](IO)$ & $2.42$ & $2.31-2.51$& $2.34-2.47$& $2.38-2.46$\\
$\theta_{12}/^\circ$ & $34.5$ & $31.5-38.0$& $32.2-36.8$& $33.5-35.7$\\
$\theta_{23}/^\circ (NO)$ & $47.7$ & $41.8-50.7$& $43.1-49.8$& $46-48.9$\\
$\theta_{23}/^\circ (IO)$ & $47.9$ & $42.2-50.7$& $44.5-48.9$& $46.2-48.9$\\
$\theta_{13}/^\circ (NO)$ & $8.45$ & $8-8.9$& $8.2-8.8$& $8.31-8.61$\\
$\theta_{13}/^\circ (IO)$ & $8.53$ & $8.1-9$& $8.3-8.8$& $8.38-8.67$\\
$\delta/^\circ (NO)$ & $218$ & $157-349$& $182-315$& $191-256$\\
$\delta/^\circ (IO)$ & $281$ & $202-349$ & $229-328$ & $254-304$ \\
\hline
\end{tabular}
\end{center}
\end{table}
\subsubsection{Majorana Yukawa Coupling for $3\sigma$ bounds of neutrino oscillation data}
We now estimate the $m_\nu$ matrix for the normally
ordered (NO) case. For this purpose we take the mass of the lightest neutrino as $m_1=0.00127$ eV. 
Then using the $3\sigma$ ranges of solar and atmospheric mass squared differences for NO case 
,as mentioned in the Table \ref{osc}, the other two neutrino mass eigenvalues are 
calculated. Obviously we get a range of values of $m_2$ and $m_3$.
Plugging in these mass eigenvalues along with all possible
combinations and the mixing angles within the $3\sigma$ bound
in eq.(\ref{mnu}) we obtain large number of sets of $m_\nu$
matrix. Thus we also get  respective bounds on the elements of 
the $m_\nu$ matrix (or equivalently on the Yukawa coupling matrix $Y$) corresponding to the $3\sigma$ oscillation constraints.
As mentioned earlier we use single set of randomly chosen Majorana
phases while the  Dirac CP phase is chosen close to its central value.
The effective light neutrino mass matrix $m_\nu$ and the coupling matrix $Y$ are connected through the induced VEV
$v_l$ which is obtained by assuming the dimensionful coupling $\mu_\Delta \sim M_\Delta$ where $M_\Delta=10^{12}$ GeV
and the electroweak VEV is $246$ GeV. With these considerations we estimate the $3\sigma$ bound on the elements of $Y$ matrix
 and present them in Table \ref{Y_no}.  
\begin{table}[!h]
\caption{Numerical values of the moduli ($|Y_{ij}|$) and phases ($\phi_{ij}$) ($i,j=1,2,3$) of Yukawa coupling matrix
for normally ordered (NO) light neutrino masses 
corresponding to $3\sigma$ global fit of neutrino oscillation data. Lightest neutrino mass eigenvalue is kept fixed at $m_1=0.00127$ eV for the sake of simplicity. 
Randomly chosen Majorana phases
$\alpha_M=74.84^\circ,\beta_M=112.85^\circ$ and the central value of
the Dirac phase $\delta=218^\circ$ have been used.}
\begin{center}
\begin{tabular}{ |c|c|c|c|c|c| } 
\hline
 $|Y_{11}|$  & $|Y_{12}|$   & $|Y_{13}|$ & $|Y_{22}|$ & $|Y_{23}|$ & $|Y_{33}|$ \\ 
   &  &  &  &  & \\ \hline
  $(1.74-3.95)\times$&  $(1.13-1.44)\times$  & $(4.09-6.71)\times$& $(3.20-4.67)\times$ & $(4.07-4.35)\times$& $(3.05-4.5)\times$\\ 
    $10^{-5}$ & $10^{-4}$  & $10^{-5}$  & $10^{-4}$& $10^{-4}$& $10^{-4}$\\ \hline\hline
  $\phi_{11}$  & $\phi_{12}$  & $\phi_{13}$ & $\phi_{22}$& $\phi_{23}$& $\phi_{33}$ \\ 
     (deg.)& (deg.)  & (deg.) & (deg.) & (deg.) & (deg.)\\ \hline
  $(-65.24)-$  & $(-48.50)-$   & $(-17.48)-$& $4.67-$& $(-6.81)-$& $3.77-$  \\ 
    $(-61.73)$& $(-44.22)$ & $8.27$ &$10.6$ & $(-5.34)$& $10.0$\\ \hline
\end{tabular}
\label{Y_no}
\end{center}
\end{table}
In the inverted mass ordering the smallest mass eigenvalue is $m_3$ which is set to be equal to $0.00127$ eV. The other two
eigenvalues are calculated using the $3\sigma$ limit of the solar and atmospheric mass squared differences. 
In this case also we are able to put a bound on the modulus and phase of the Yukawa coupling matrix 
following the same procedure as done in the case of NO. The constrained parameters ($|Y_{ij}|,\phi_{ij}$)  
for inverted mass ordering are given in Table \ref{Y_io}. 
\begin{table}[!h]
\caption{Numerical values of the moduli ($|Y_{ij}|$) and phases
  ($\phi_{ij}$) ($i,j=1,2,3$) of Yukawa coupling matrix $Y$
for invertedly ordered (IO) light neutrino masses 
corresponding to $3\sigma$ global fit of neutrino oscillation data. Lightest neutrino mass eigenvalue is kept fixed at $m_3=0.00127$ eV. 
Phase angles used are the same as in Table \ref{Y_no}.}
\begin{center}
\begin{tabular}{ |c|c|c|c|c|c| } 
\hline
 $|Y_{11}|$  & $|Y_{12}|$   & $|Y_{13}|$ & $|Y_{22}|$ & $|Y_{23}|$ & $|Y_{33}|$ \\ 
   &  &  &  &  & \\ \hline
  $(4.38-5.3)\times$&  $(4.29-5.5)\times$  & $(3.55-4.87)\times$& $(8.83-23.5)\times$ & $(2.13-2.89)\times$& $(2.84-4.0)\times$\\ 
    $10^{-4}$ & $10^{-4}$  & $10^{-4}$  & $10^{-5}$& $10^{-4}$& $10^{-4}$\\ \hline\hline
  $\phi_{11}$  & $\phi_{12}$  & $\phi_{13}$ & $\phi_{22}$& $\phi_{23}$& $\phi_{33}$ \\ 
     (deg.)& (deg.)  & (deg.) & (deg.) & (deg.) & (deg.)\\ \hline
  $52.96$  & $(-6.51)-$   & $0.5$& $(-60)-$& $(-69.16)-$& $(-78.89)-$  \\ 
    $68.35$& $(-3.16)$ & $4.5$ &$(-32.31)$ & $(-51.39)$& $(-61.92)$\\ \hline
\end{tabular}
\label{Y_io}
\end{center}
\end{table}
\vskip .3cm
As we have taken a most general complex symmetric structure of the $m_\nu$ matrix (or in other words the Yukawa coupling matrix $Y$)
without imposing any kinds specific flavor symmetry, it doesn't have any definite prediction of the Dirac CP violating phase $\delta$.
Any value of $\delta$ in the given $3\sigma$ range can be accommodated. In this regard few remarks about the present experimental status
of the Dirac CP phase are in order. The recent global analysis of oscillation data done in Ref\cite{Salas:2017} has made it clear that
value of the Dirac CP phase $\delta=\pi/2$ is more or less ruled out. In Normal mass ordering (NO) $\delta=\pi/2$ is disfavored at more than $4\sigma$
confidence level whereas for inverted mass ordering (IO) it is more stringent, where $\delta=\pi/2$ is ruled out at more than $6\sigma$. The best fit value 
of $\delta$ in NO and IO are near $1.2\pi$ and $1.5\pi$, respectively. For the sake of simplicity we work with only the best fit values.
We have also estimated the highest and lowest values of the CP violating measure, the Jarlskog invariant $(J_{CP}=-s_{12}c_{12}s_{13}c_{13}^2s_{23}c_{23} \sin \delta)$
for both the mass orderings. For NO: $J_{CP}=0.0175-0.0212$, for IO :$J_{CP}=0.0302-0.0365$  when $\delta$ is kept fixed at its best fit value whereas all other observables 
are varying in their respective $3\sigma$ ranges.

\subsubsection{Majorana Yukawa Coupling for $1\sigma$ Bounds of Neutrino Oscillation Data}
\begin{figure}[!h]
\begin{center}
\includegraphics[width=6cm,height=6cm,angle=270]{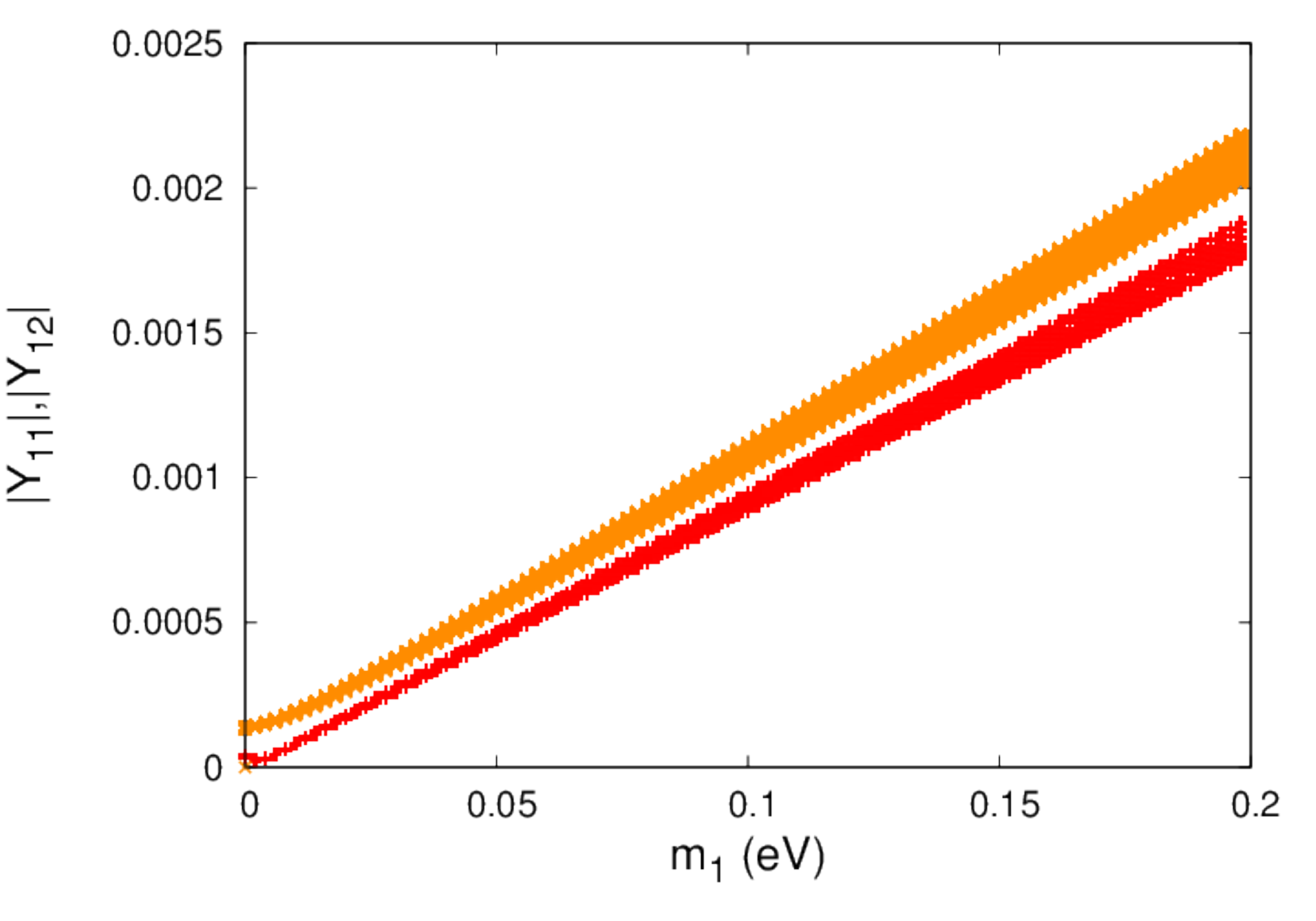}
\hspace{1cm}
\includegraphics[width=6cm,height=6cm,angle=270]{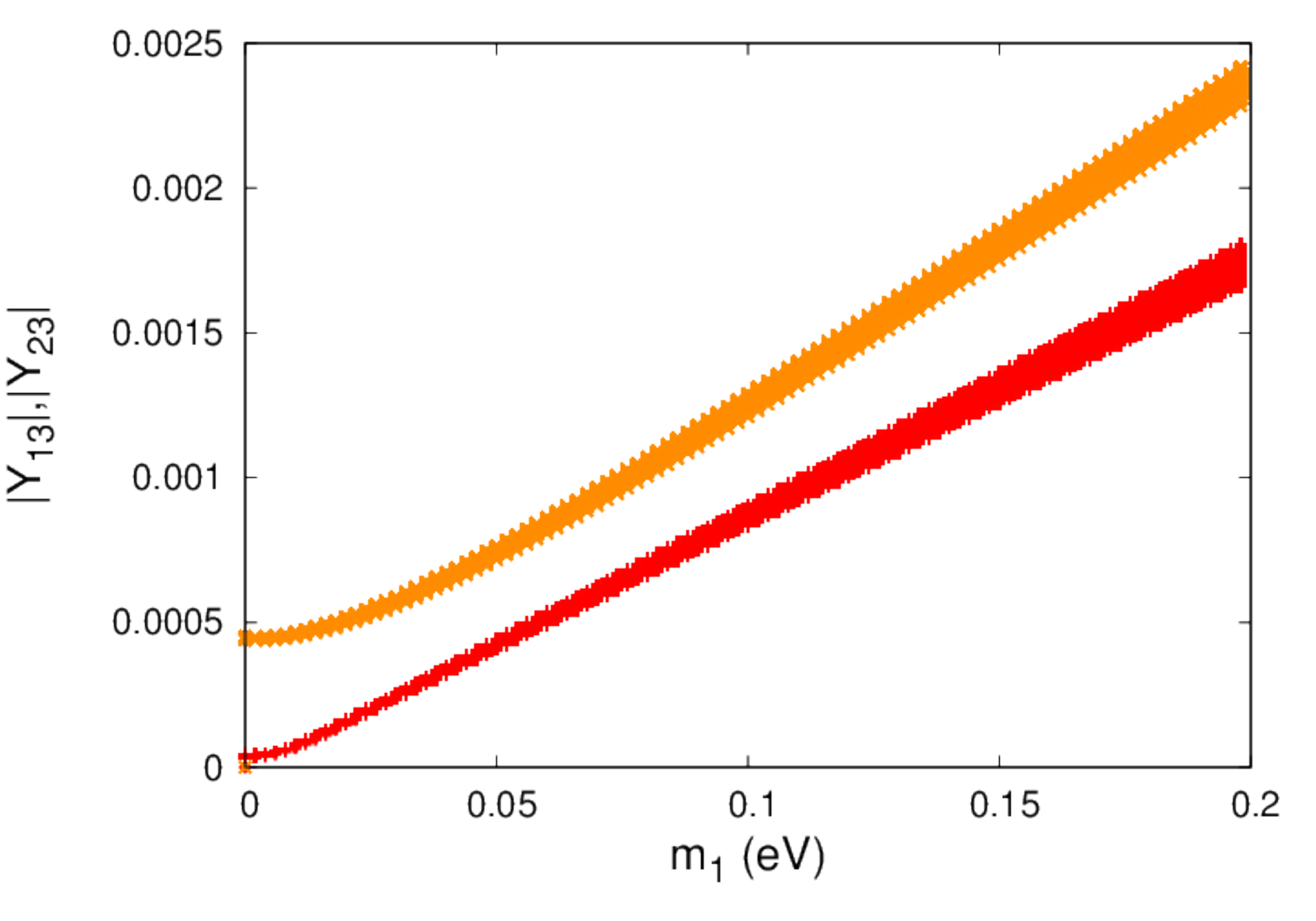}
\caption{Determination of moduli of $Y$ matrix elements within
  $1\sigma$ uncertainty of oscillation data as a function of
 lightest neutrino mass
  eigenvalues $m_1$. Phase angles used  are randomly
  chosen Majorana phases $\alpha_M=37.91^\circ,\beta_M=157.91^\circ$
and central value of the Dirac phase $\delta=216^\circ$. In the left panel red and yellow regions denote $1\sigma$ allowed values $|Y_{11}|$ and $|Y_{12}|$, respectively.
In the right panel red patch gives values of $|Y_{13}|$  whereas
yellow region denotes the same for $|Y_{23}|$ within the same
uncertainty of the oscillation data.  }
\label{y_mod1}
\end{center}
\end{figure}

\begin{figure}[!h]
\begin{center}
\includegraphics[width=6cm,height=6cm,angle=270]{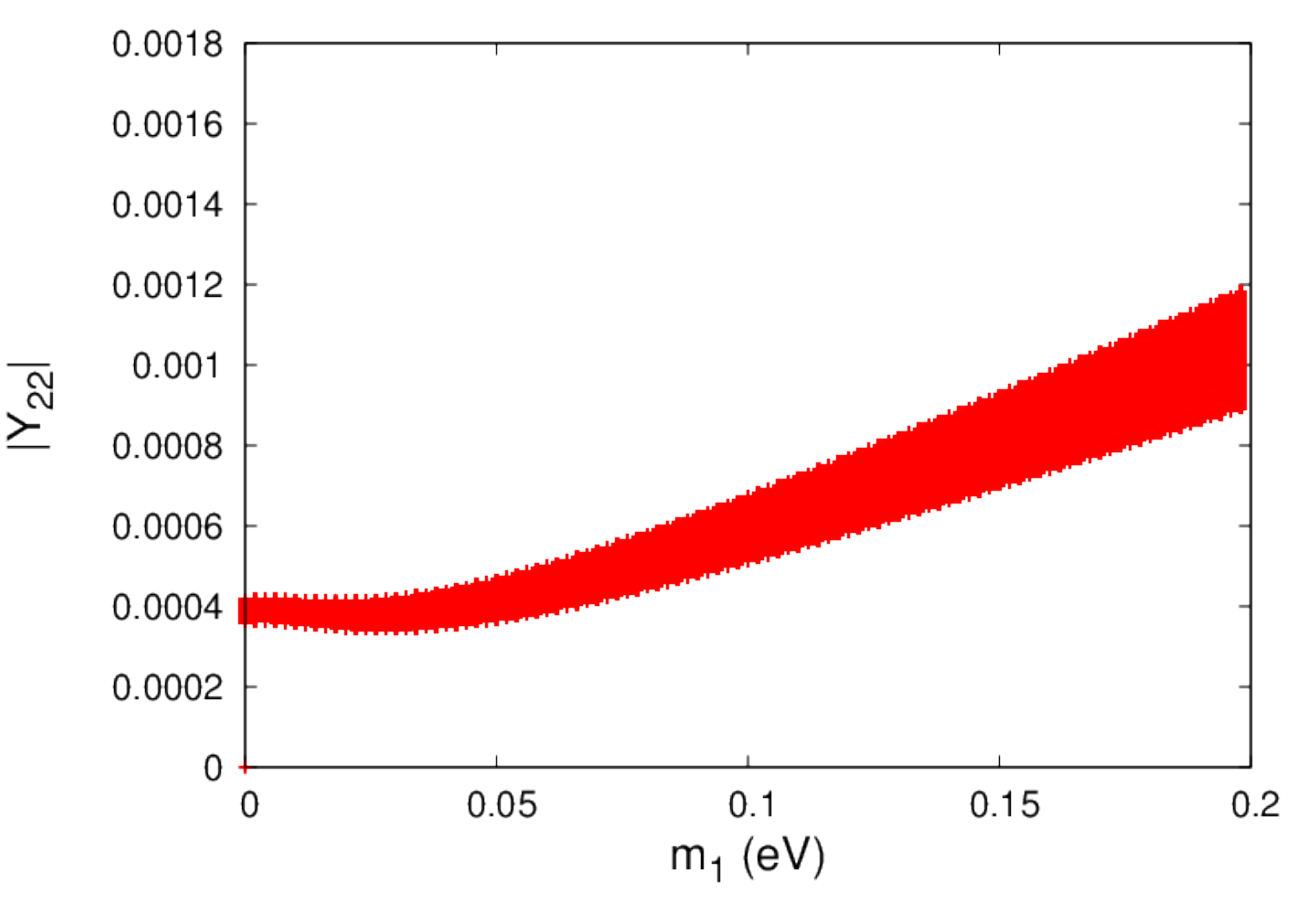}
\hspace{1cm}
\includegraphics[width=6cm,height=6cm,angle=270]{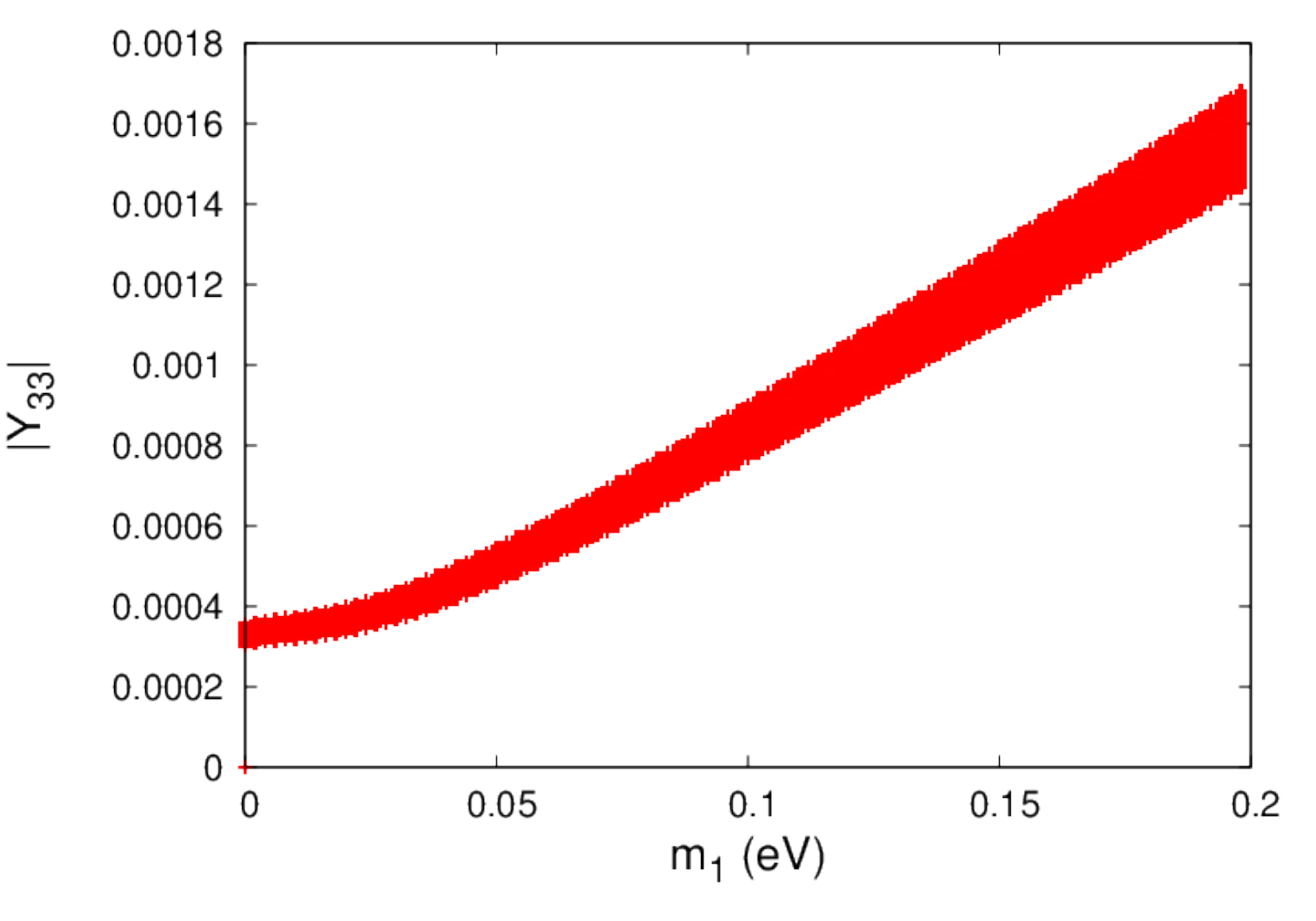}
\caption{Determination of moduli of $Y$ matrix elements within
  $1\sigma$ uncertainty of oscillation data as a function of lightest
  neutrino mass eigenvalues $m_1$ as shown in the left panel for
  $|Y_{22}|$, and in the right panel for  $|Y_{33}|$. Phase angles used are the same as in Fig. \ref{y_mod1}. }
\label{y_mod2}
\end{center}
\end{figure}

\begin{figure}[!h]
\begin{center}
\includegraphics[width=4.7cm,height=4.7cm,angle=270]{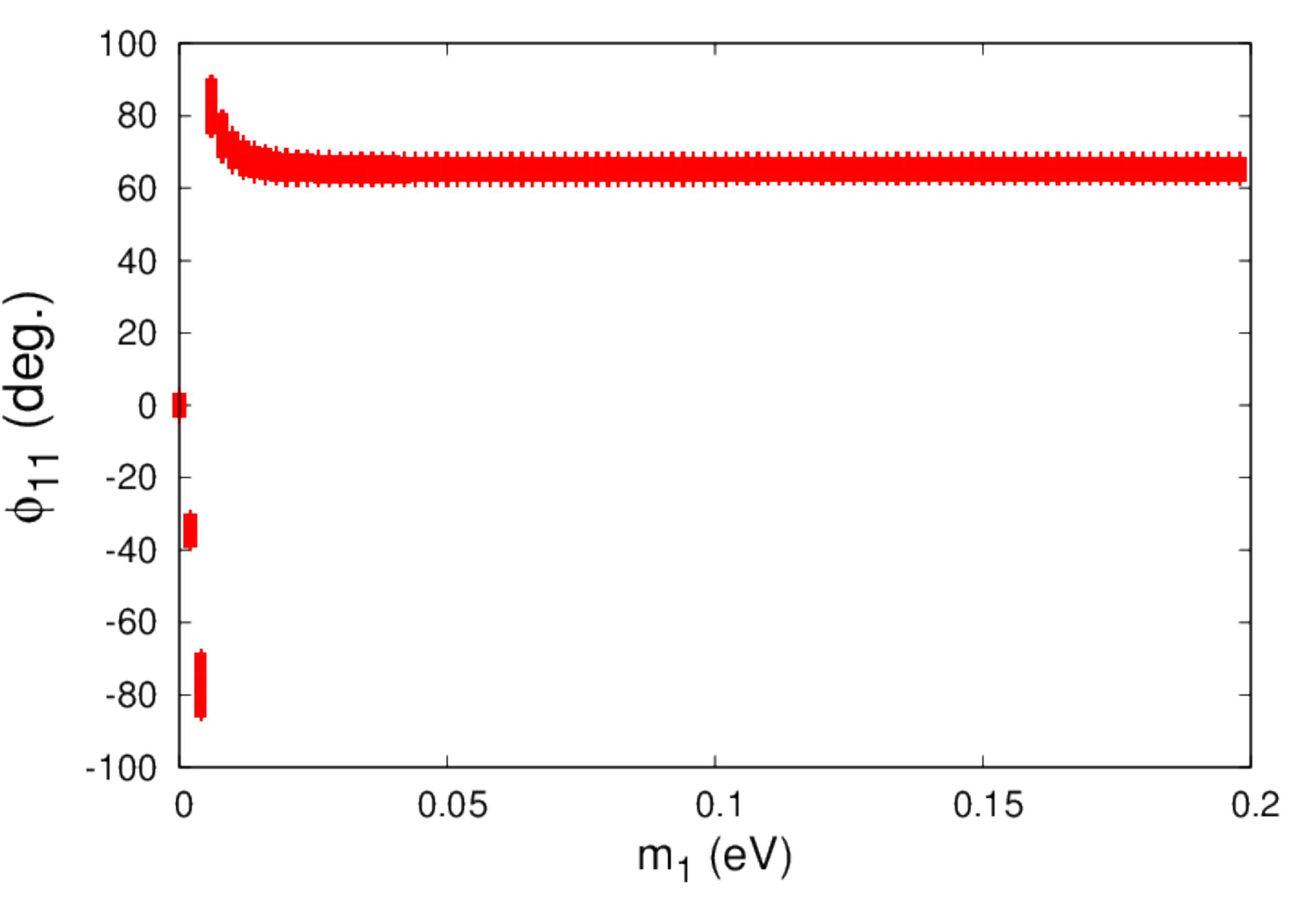}
\hspace{.5cm}
\includegraphics[width=4.7cm,height=4.7cm,angle=270]{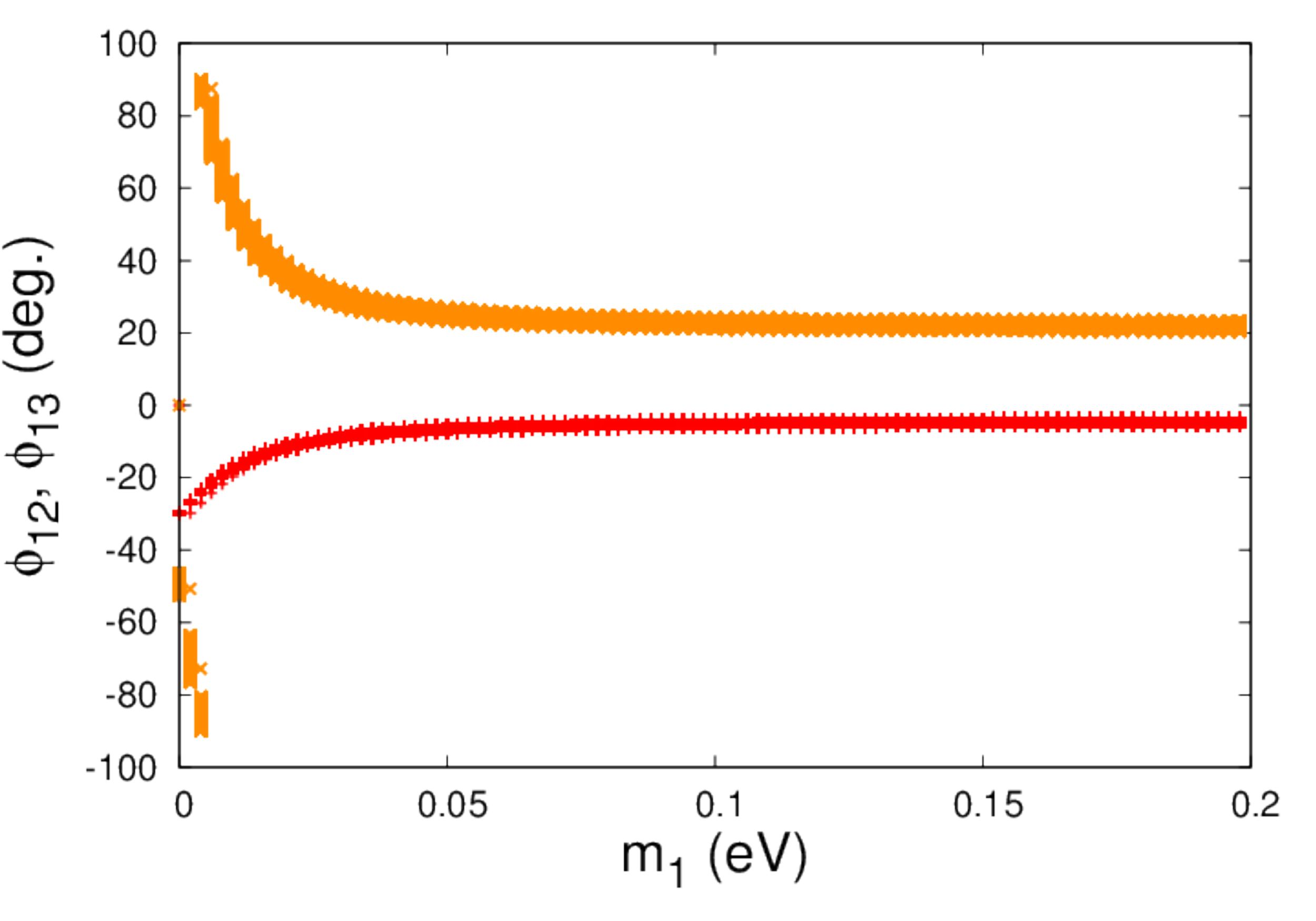}
\hspace{.5cm}
\includegraphics[width=4.7cm,height=4.7cm,angle=270]{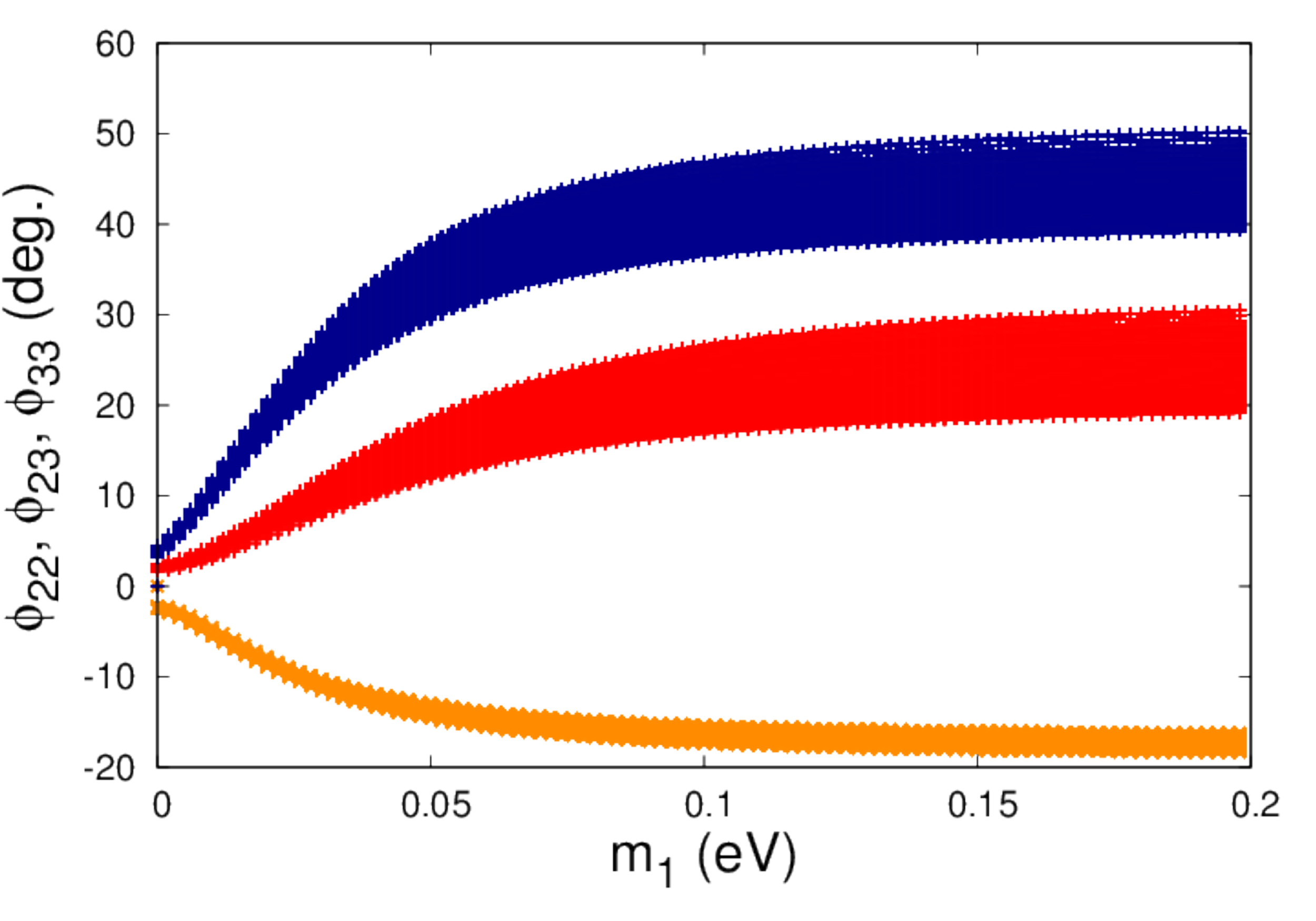}
\caption{Determination of phases of $Y$ matrix elements as a function
  of lightest neutrino mass
  eigenvalues $m_1$ within $1\sigma$ allowed uncertainty of oscillation
  data.  Phase angles used are the same as in Fig. \ref{y_mod1}.
 In the extreme left panel red region denotes  values of $\phi_{11}$. In the middle panel red and yellow
regions denote  allowed values $\phi_{12}$ and $\phi_{13}$, respectively.
In the extreme right panel red, yellow and blue patches give allowed
values of $\phi_{22},\phi_{23}$ and $\phi_{33}$, respectively. }
\label{phase1}
\end{center}
\end{figure}

\begin{figure}[!h]
\begin{center}
\includegraphics[width=4.7cm,height=4.7cm,angle=270]{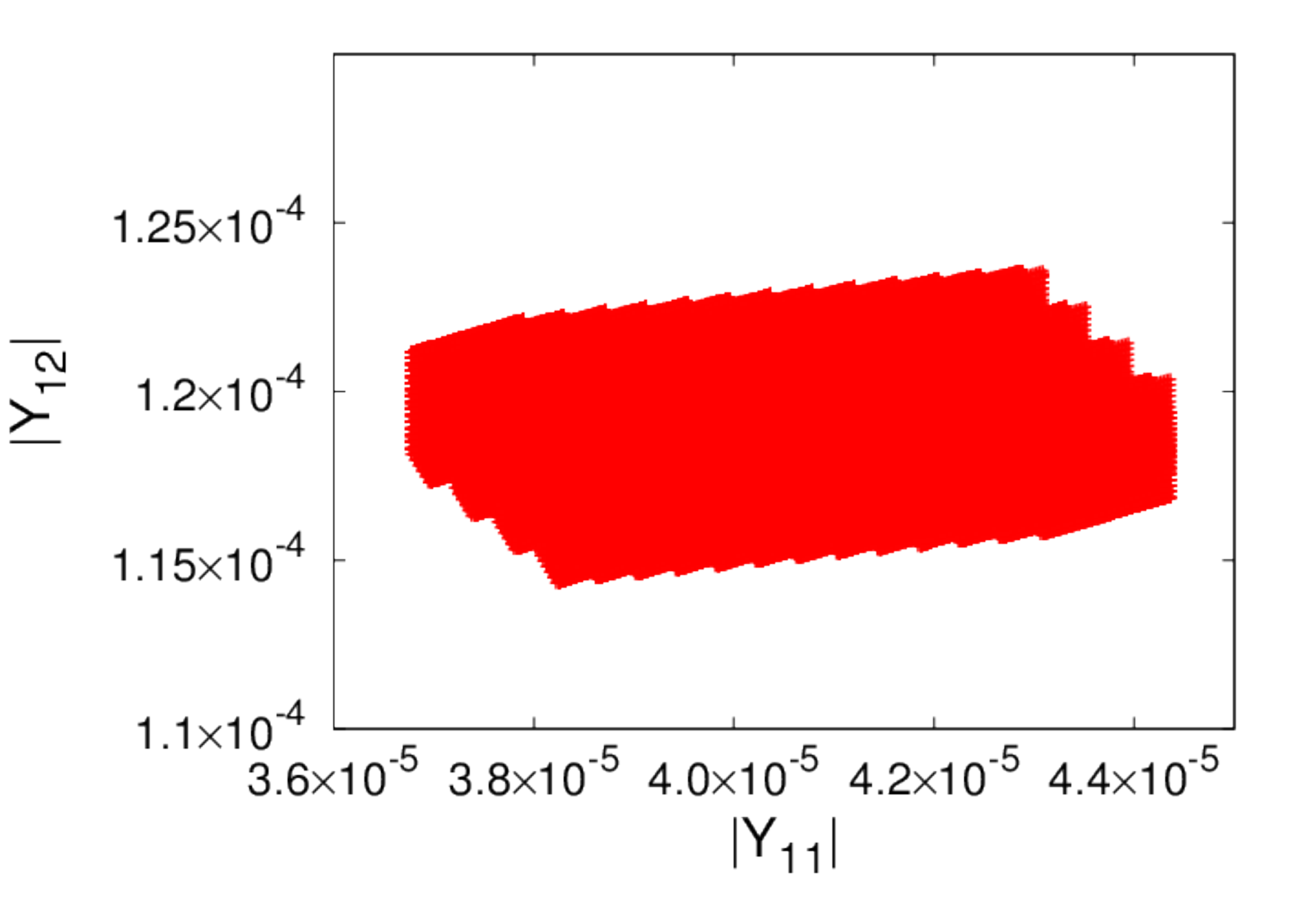}
\hspace{.5cm}
\includegraphics[width=4.7cm,height=4.7cm,angle=270]{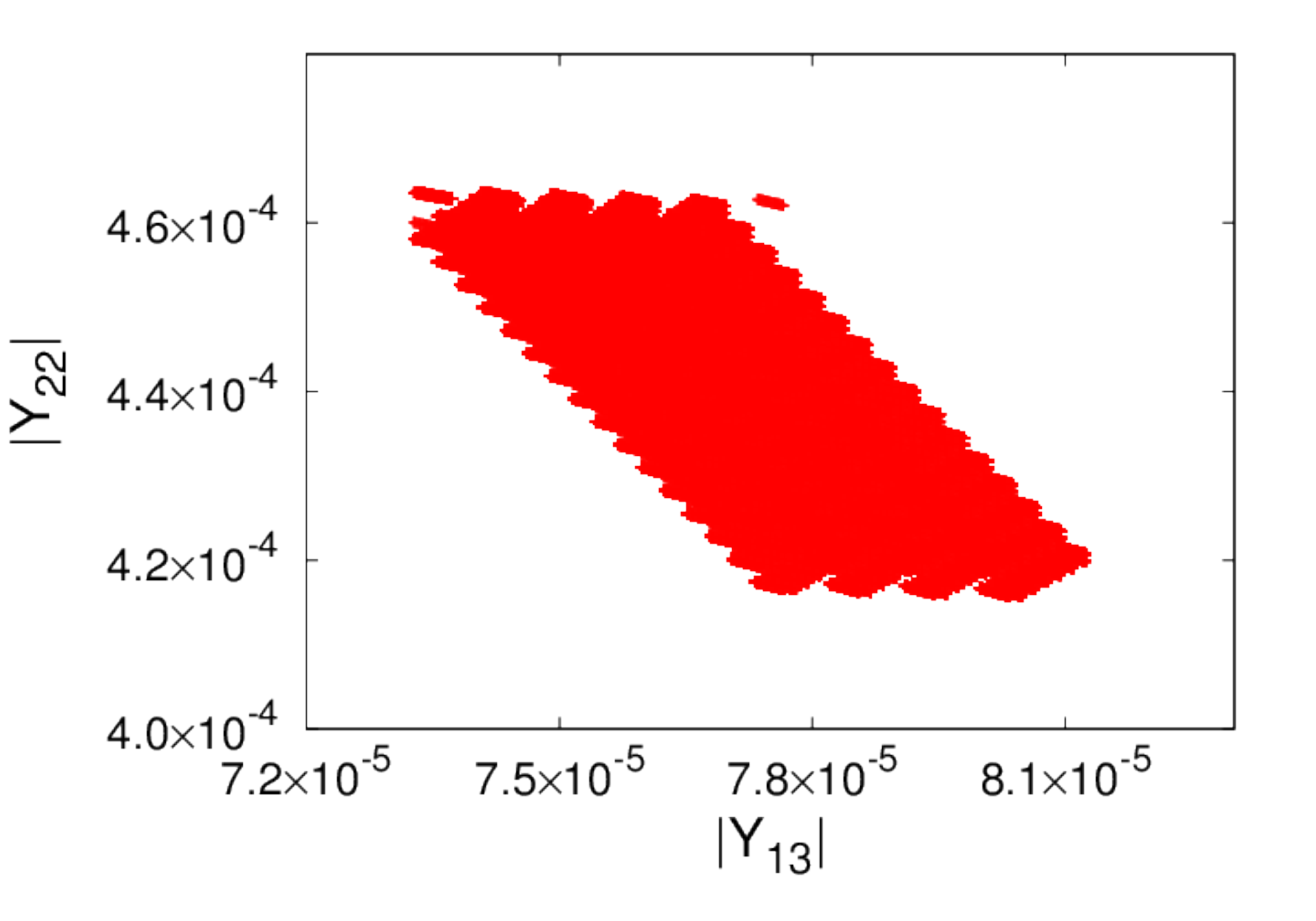}
\hspace{.5cm}
\includegraphics[width=4.7cm,height=4.7cm,angle=270]{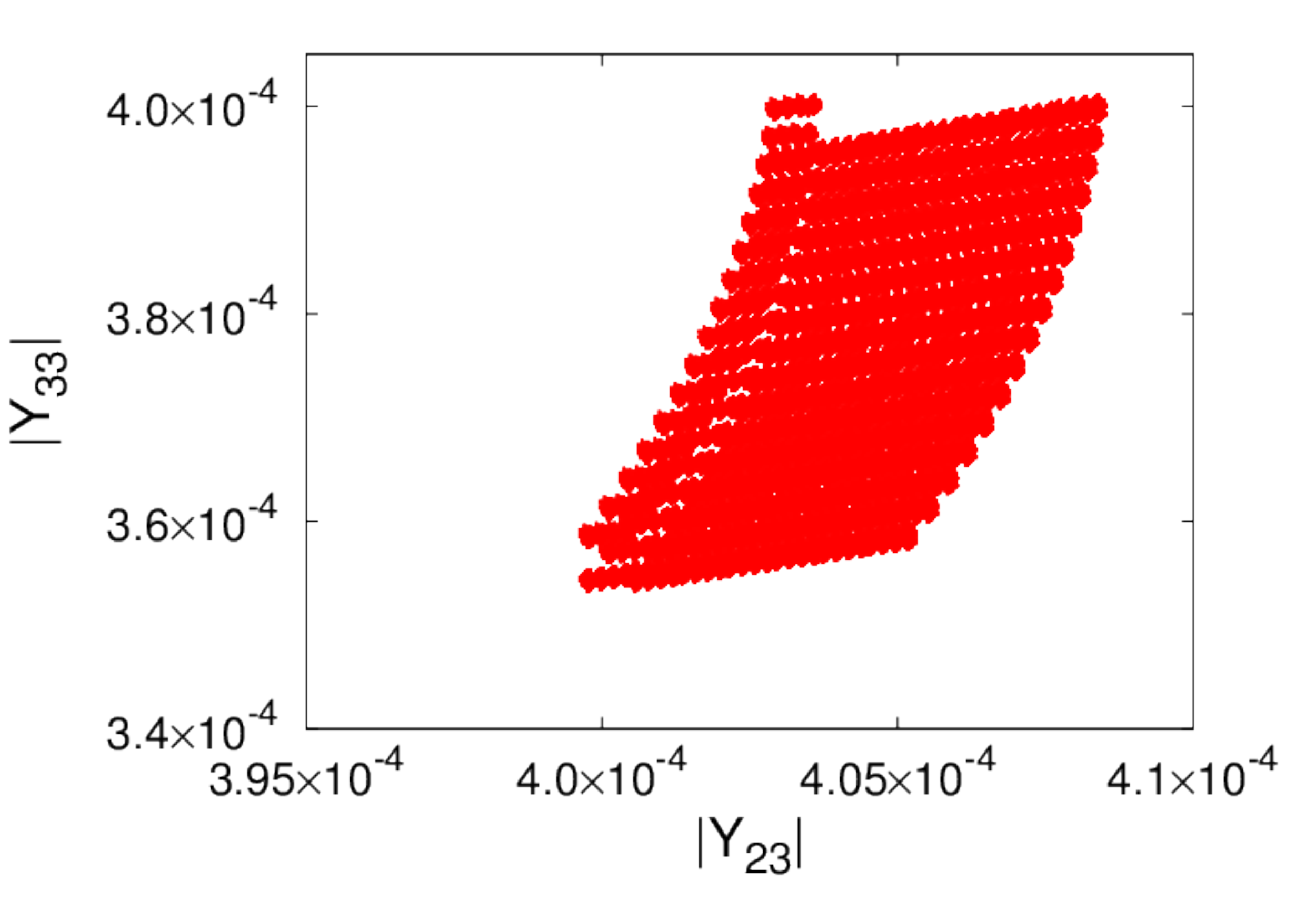}
\caption{Determination  of moduli of $Y$ matrix elements for a fixed value of lightest mass eigenvalue $m_1=0.00127$ eV. 
Phase angles used for computaion are $\alpha_M=124.37^{\circ},
\beta_M=86.27^{\circ}$ (randomly chosen) and $\delta=216^{\circ}$. Neutrino oscillation observables are varied within
$1\sigma$ range. Left panel: variation of $|Y_{11}|$ with $|Y_{12}|$,
middle panel: $|Y_{13}|$ vs $|Y_{22}|$, right panel:  $|Y_{23}|$ vs $|Y_{33}|$.}
\label{y_mod11}
\end{center}
\end{figure}

\begin{figure}[!h]
\begin{center}
\includegraphics[width=4.7cm,height=4.7cm,angle=270]{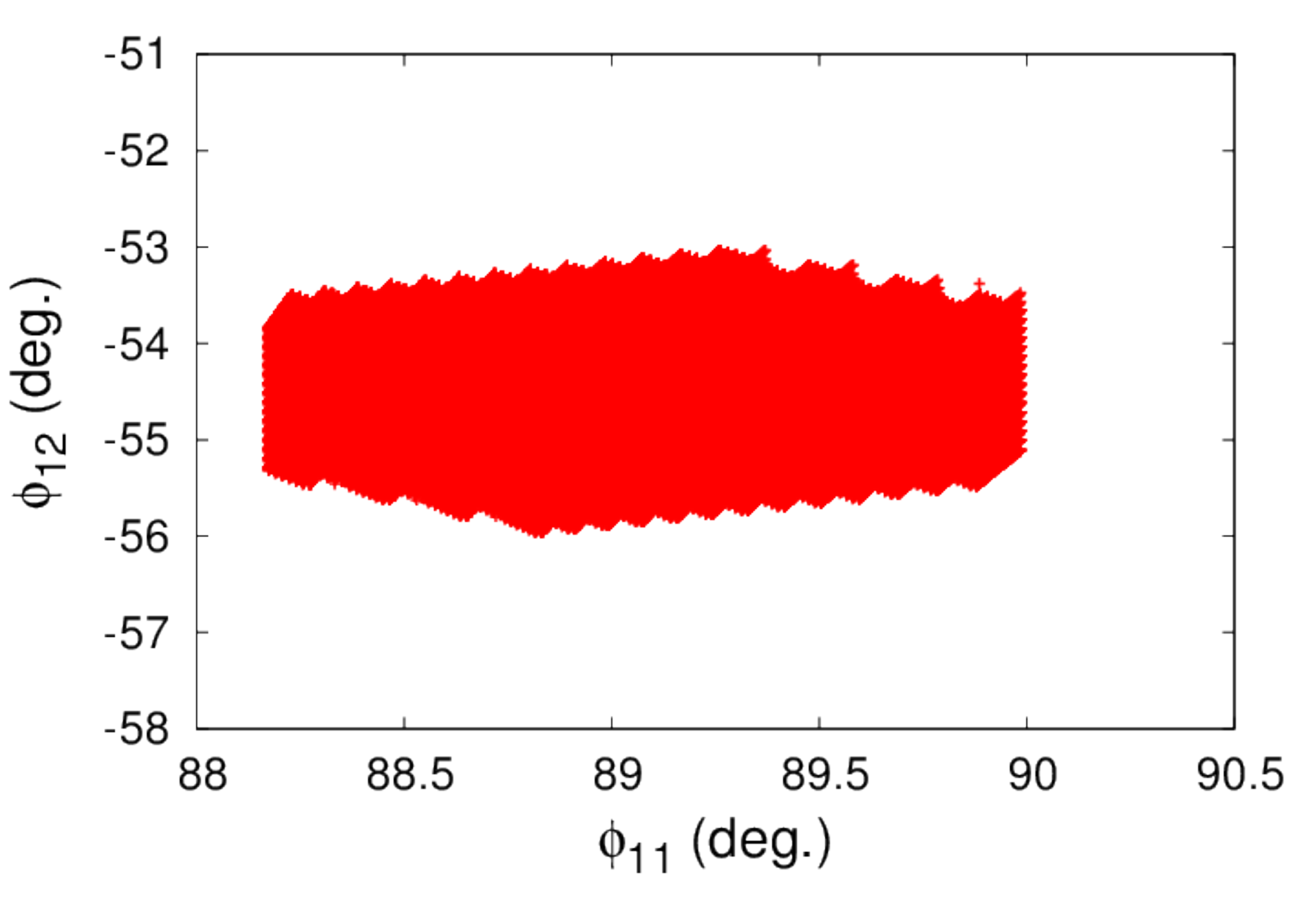}
\hspace{.5cm}
\includegraphics[width=4.7cm,height=4.7cm,angle=270]{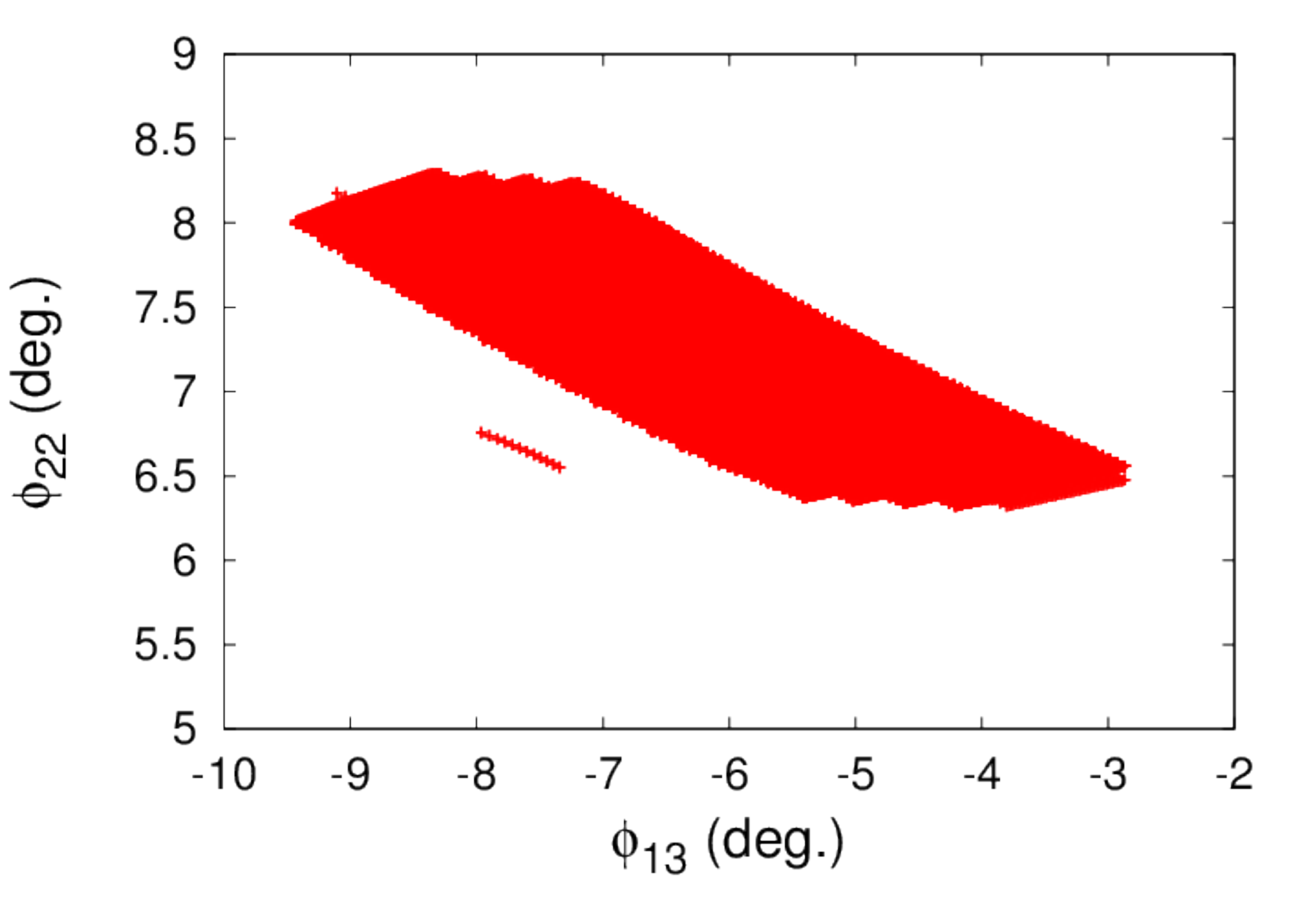}
\hspace{.5cm}
\includegraphics[width=4.7cm,height=4.7cm,angle=270]{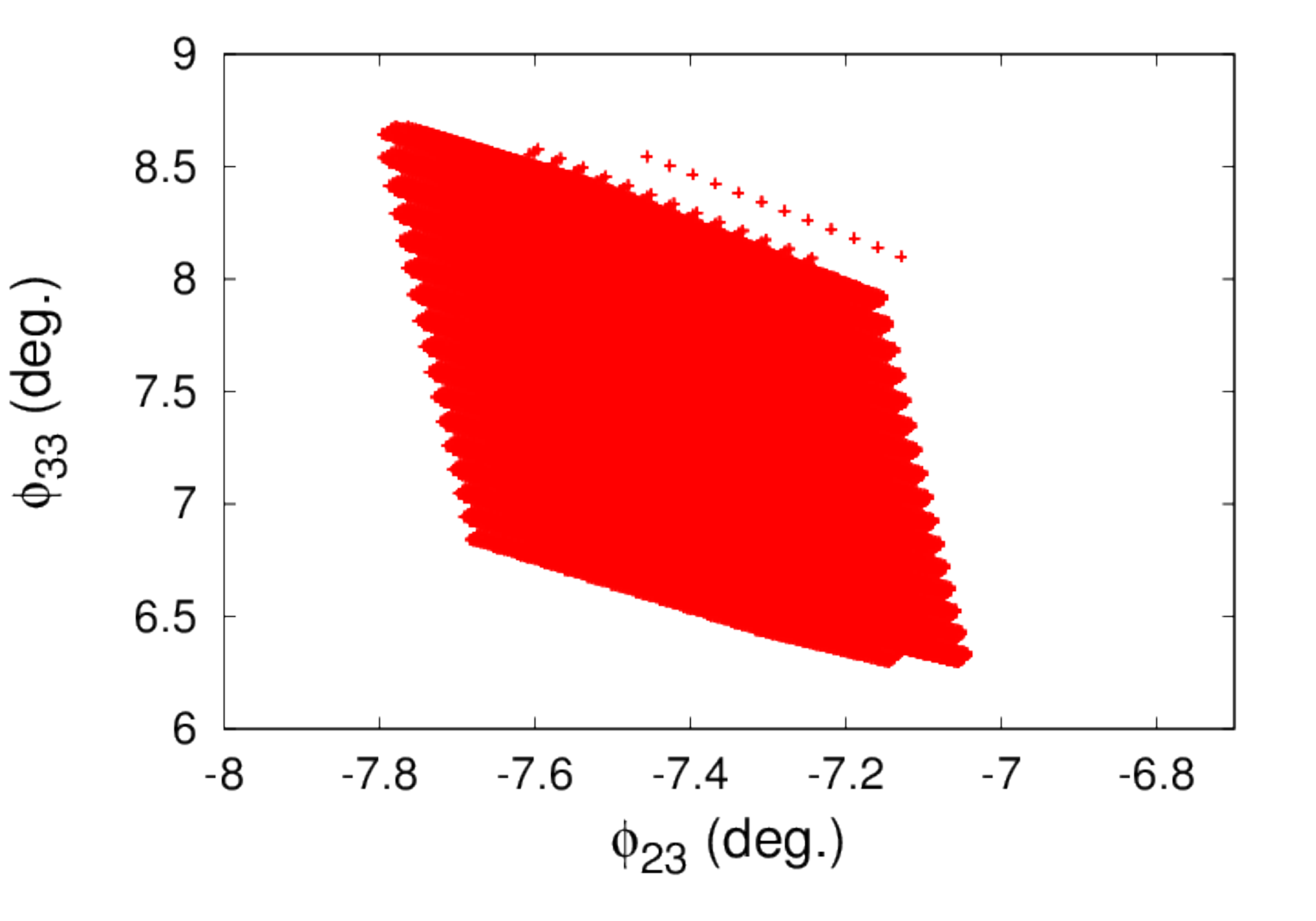}
\caption{Determination  of phases of $Y$ matrix elements for fixed value of
  lightest mass eigenvalue $m_1=0.001$ eV. Phase angles used for
  computaion are the same as in Fig. \ref{y_mod11}.
 Neutrino oscillation observables are varied within $1\sigma$ range; left panel: variation of $\phi_{11}$ with $\phi_{12}$,
middle panel: $\phi_{13}$ vs $\phi_{22}$, right panel:  $\phi_{23}$ vs. $\phi_{33}$.}
\label{phase22}
\end{center}
\end{figure}

Here we follow exactly same methodology as the previous case, however the numerical calculations are done with $1\sigma$ ranges of oscillation 
data instead of $3\sigma$ range. Here we are exploring the normally ordered case only. Unlike the previous case the lightest neutrino mass eigenvalue
$m_1$ isn't kept fixed, it is varied over a range of $(0-0.2)$ eV and the corresponding variations of the modulus and phase of Majorana Yukawa
couplings are depicted in Figs.\ref{y_mod1}, \ref{y_mod2}, \ref{phase1}. The $1\sigma$ allowed range of those quantities for a fixed $m_1$ are also shown 
in Figs.\ref{y_mod11}, \ref{phase22}. 


\section{Gauge Coupling Unification in the Scalar Extended SU(5)}\label{sec:unif} 
\subsection{ Lower Bound on the Scalar Triplet Mass}
Exercising utmost economy in populating the grand desert, it was noted that
 the presence of the scalar component $\kappa
(3,0,8) \subset {75}_H$ at an intermediate mass $\simeq 10^{10}$ GeV
could achieve  gauge coupling
unification at $M_{GUT} \simeq 10^{15}$ GeV
\cite{kynshi-mkp:1993} but no  neutrino oscillation data was available
at that time.  
Using the most recent electroweak precision data \cite{PDG:2012,PDG:2014,PDG:2016}, in this work we find that  this intermediate scalar mass is now reduced by one order, $M_{\kappa}=10^{9.23}$ GeV. Similarly the GUT scale is now determined with high precision including all possible theoretical and experimental uncertainties. Noting the result of this work as discussed in Sec. \ref{sec:numass} that type-II seesaw realisation of neutrino mass needs $M_{\Delta}$ substantially lower than the GUT scale, leads to the natural apprehension that the presence $\Delta_L (3,-1,1)$ at intermediate mass scale would destroy precision unification achieved by $\kappa (3,0,8)$. This apprehension is logically founded on the basis that nonvanishing contributions to the $SU(2)_L$ and $U(1)_Y$ beta functions would misalign the finestructure constants $\alpha_{2L}(\mu)$ and $\alpha_Y(\mu)$ from the $\kappa(3,0,8)$ realised unification paths substantially for all mass scales $\mu > M_{\Delta}$.
 
We prevent any such deviation from the  $\kappa (3,0,8)$-realisation of precision coupling unification by assuming all the components of ${15}_H\subset SU(5)$ to have the identical degenerate mass $M_{\Delta}$ which is bounded in the following manner

\begin{equation}           
M_{\kappa} \le \left (M_{\Delta} =M_{{15}_H}\right) \le M_{\rm GUT}. \label{eq:bound}
\end{equation}
Thus, in order to safeguard precision unification, it is essential
that $M_{\Delta}=M_{{15}_H} \ge M_{\kappa}$ in the present
scalar-extended SU(5) model \footnote{ The upper limit is due to our
  observation that type-II seesaw scale is lower than the GUT scale
  although, strictly speaking,  $M_{\Delta} =M_{{15}_H} >  M_{\rm
    GUT}$ is possible if type-II seesaw contribution to neutrino mass
  is ignored.}.

Thus  type-II seesaw realisation of neutrino mass and precision unification in SU(5) needs the additional scalar representations ${15}_H$ and ${75}_H$.

\subsection{RG Solutions to Mass Scales}
For realistic unification of gauge couplings we use one loop equations \cite{GQW} supplemented by top-quark threshold effects \cite{Langacker:1993} and two-loop corrections \cite{Jones:1982}

\begin{eqnarray}
\mu\frac{{{\partial g}}_i(\mu)}{{\partial \mu}}&=& \frac{a_i}{16 \pi^2} g_i^3 \nonumber\\
&+&\frac{1}{(16\pi^2)^2}\left[\Sigma_{j}b_{ij}g_i^3 g_j^2-\kappa_iy_{\rm top}^2]\right. \label{rges-diff}
\end{eqnarray}

 In the range of mass scale $\mu=M_Z-M_{\rm U}$ we include
 top-quark Yukawa coupling  ($y_{top}$) contribution at the two-loop level with
the coefficients in eq.(\ref{rges-diff}) $\kappa_{1Y}=17/10, \kappa_{2L}=3/2,\kappa_{3C}=2$ and the RG
 evolution equation \cite{Langacker:1993}
\be
\mu\frac{\partial {y_{top}}}{\partial \mu}=
\frac{y_{top}}{16\pi^2}(\frac{9}{2}y_{top}^2-\frac{17}{20}g_{1Y}^2-\frac{9}{4}g_{2L}^2-8g_{3C}^2). \label{rge-ytop}
\ee   
 The beta function coefficients in three different mass ranges
 $\mu=M_Z \to M_{\kappa}$, $\mu= M_{\sigma}-M_{\Delta}$, and $\mu=M_{\Delta}-M_U$ are 
\par\noindent{\large\bf {\underline {$\mu = M_{Z} \to M_{\kappa}:$}}}\\ 
\begin{equation}
 a_Y=\frac{41}{10},\,\, a_{2L}=-\frac{19}{6},\,\, a_{3C}=-7, \label{eq:smai}
\end{equation}
\par\noindent{\large\bf {\underline {$\mu=  M_{\kappa}\to
      M_{\Delta}:$}}}\\
\begin{equation}
 a_Y^{\prime}=\frac{41}{10},\,\,
 a_{2L}^{\prime}=-\frac{1}{2},\,\, a_{3C}^{\prime}= -{11 \over 2},\label{eq:kapai}
\end{equation}
\par\noindent{\large\bf {\underline {$\mu= M_{\Delta}\to M_{U}:$}}}\\ 
\begin{equation}
 a_Y^{\prime\prime}=\frac{79}{15},\,\, a_{2L}^{\prime\prime}=\frac{2}{3},\,\,
  a_{3C}^{\prime\prime}=-{13 \over 3}
 .\label{eq:delai}
\end{equation}
We have used the most recent  electroweak
precision data \cite{PDG:2016}
\ba   
\alpha_S(M_Z)&=& 0.1182\pm 0.0005, \, \nonumber\\
\sin^2\theta_W (M_Z)&=&0.23129\pm 0.00005,\,\nonumber\\
\alpha^{-1}(M_Z)&=& 127.94 \pm 0.02.\, \label{eq:inputpara}
\ea

Using RGEs and  the combinations ${1 \over \alpha(M_Z)}-{8 \over 3}{1 \over \alpha_{2L}(M_Z)}$ and  ~~ ${1 \over \alpha(M_Z)}-{8 \over 3}{1 \over \alpha_{3C}(M_Z)}$, we have derived analytic formulas for the unification scale and intermediate scale( $M_ {\kappa}$) treating $SU(2)_{L}$ triplet scalar scale($M_{\Delta}$) constant as: 

\begin{align}
\ln {M^0_U \over M_Z} &= \frac{2\pi}{187\alpha}\left(7-\frac{80\alpha}
{3\alpha_{3C}}+8s_W^2 \right)+\Delta_U \nonumber\\
\ln {M^0_{\kappa}  \over M_Z}& = \frac{12\pi}{187\alpha}\left(5+\frac{23\alpha}
{3\alpha_{3C}}-21s_W^2 \right)+\Delta_{\kappa}  \nonumber\\
{1 \over \alpha_{G^0}}& = {3 \over 8\alpha}+{1 \over 187\alpha}\left({347 \over 8}
 +{466\alpha \over 3\alpha_{3C}}-271s_W^2\right)+\Delta_{\alpha_G} \label{eq:guform} 
\end{align}

where $s_W^2=\sin^2\theta_W (M_Z)$ and the first term in the above eq.(\ref{eq:guform}) represent one loop contributions. The terms $\Delta_I^i$, $i=U,~\kappa, ~ \alpha_G$ denote the threshold corrections due to unification scale($M_U$), intermediate scale ($M_{\kappa}$) and GUT fine structure constant({$1 \over \alpha_G$}).


Excellent unification of gauge couplings is found for
\ba
M^0_U  &=& 10^{15.2+0.0312 }\, {\rm GeV},  \nonumber\\
M^0_{\kappa} &=& 10^{9.23}\, {\rm GeV}, \nonumber\\
\alpha^{-1}_{G_0} &=& 41.79 \, \label{eq:massscaleseta}  
\ea
where the number $0.0312$ in the exponent is due to GUT scale matching
of inverse finestructure constant that is present even if all
superheavy masses are exactly at $\mu=M^0_U$ \cite{Hall:1981,mkp:1987}.

\subsection{Effects of ${15}_H$ on Unification}
It is well known that when a complete GUT representation is
superimposed on an already realised unification pattern in  
non-SUSY GUTs \cite{mkp:2011,RNM-mkp:2011}, the GUT scale
is unchanged but the inverse finestructure constants change their
slopes and deviate from the original paths proportionately so as to
increase the unification coupling. As an example in non-SUSY
SO(10) \cite{mkp:2011,RNM-mkp:2011}, at
first a precision unification frame has been achieved with the  modification of the TeV
scale spectrum of the minimal SUSY GUT by taking out the full scalar super partner
content of the spinorial super field representation ${16}\subset SO(10)$. Then the resulting TeV scale spectrum is \cite{mkp:2011,RNM-mkp:2011}
\begin{equation}
\chi(2,-1/2, 1),\,F_{\phi}(2,1/2,1),\,F_{\chi}(2, -1/2,1),\,
F_{\sigma}(3,0,1)\,, F_b(1,0,1)\,, F_c(1,0,8) \label{eq:tevsp}
\end{equation} 
which may be recognised to be the same as the corresponding spectrum in
the split-SUSY case supplemented by the additional scalar doublet
$\chi(2,-1/2, 1)$. In eq.(\ref{eq:tevsp}) $F_i$'s represent
nonstandard fermions. Further adjustment of masses of these particles
around TeV scale has been noted to achieve degree of precision
coupling unification higher than MSSM \cite{mkp:2011}. 
After having thus achieved a precision unification, the full ${15}_H$ is superimposed at the
type-II seesaw scale $M_{\Delta}$ of the non-supersymmetrised
unification framework. Analogous to MSSM, 
this model \cite{RNM-mkp:2011} predicts a number of
 fermions as in eq.(\ref{eq:tevsp}) at the TeV scale which must be
verified experimentally at accelerator energies. \\
 
In contrast, the
present model has only the standard Higgs doublet $\phi (2,1/2,1)$ and
the WIMP DM  scalar singlet $\xi (1,0,1)$ near TeV scale as discussed below.
 Although the TeV scale DM has not been
confirmed yet by direct experiments, LUX-16 or Femi-LAT like
experiments may detect it.
Moreover, as shown below, the present model ensures vacuum
stability through this WIMP dark matter candidate whereas in
\cite{RNM-mkp:2011} the vacuum stability and  DM issues are yet to
be answered. Further, the SM coupling unification scale in \cite{RNM-mkp:2011} being close to
the SUSY GUT scale,
$M_U \sim 10^{16}$ GeV,  predicts proton lifetime nearly $60$
times larger than the current experimental limit without threshold effect which
is expected to introduce larger  uncertainty compared to the present model.  
It may be more difficult to verify this model  by ongoing proton decay
experiments. But the present model including such uncertainties is
within the experimentally accessible limits.
The origin of invariant GUT scale in the presence of ${15}_H$ in the present model is due to the invariance of the beta
function differences which is $-7/6$ in this model
\begin{equation}
\Delta a_i= (a_i^{\prime}-a_i^{\prime\prime})=-(7/6) \,\, ,(i=1,2,3).\label{eq:invar}
\end{equation}
This results in a change in the inverse GUT coupling
constant $\alpha_G^{-1}$ which occurs due to the RG predicted
modification
\begin{equation}
{1 \over \alpha_G} = {1 \over \alpha_{ G^0}}-{1 \over 561\alpha}\left({229 \over 2}
 +{134\alpha \over 3\alpha_{3C}}-350s_W^2\right)+ {7 \over 12\pi}\ln\left({M_{\Delta}\over M_Z}\right).
\end{equation}
 

Thus the  result of type-II seesaw motivated insertion of ${15}_H$ into the $\kappa -$ realised unification framework is to decrease in the inverse GUT fine structure constant while keeping mass scales same as in eq.(\ref{eq:massscaleseta})
\begin{equation}
\alpha^{-1}_G = 37.765. \label{eq:Deleff}
\end{equation}
which is a $9.6\%$ effect. It is essential to take this effect into
consideration in the top-down approach for consistency with the
precision value of the electromagnetic finestructure constant $\alpha
^{-1}(M_Z)=127.9\pm 0.01$ \cite{PDG:2016}. More important is its
visible effect on proton lifetime prediction.
It is clear from $\alpha_G^{-2}$ dependence in eq.(\ref{taup}) of
Sec.\ref{sec:plife}, eq.(\ref{eq:massscaleseta}) and
eq.(\ref{eq:Deleff}) that the this intermediate type-II seesaw scale
has a proton lifetime reduction by $19\%$ that further reduces for
lower seesaw scales,$M_{\Delta}< 10^{12}$ GeV. But the reduction
effect decreases as $M_{\Delta}$ increases such that the proton
lifetime remains unchanged for the limiting value $M_{\Delta}=M_{{15}_H}=M_{U}$.\\ 

\begin{figure}[h!]
\begin{center}
\includegraphics[height=8cm,width=8cm]{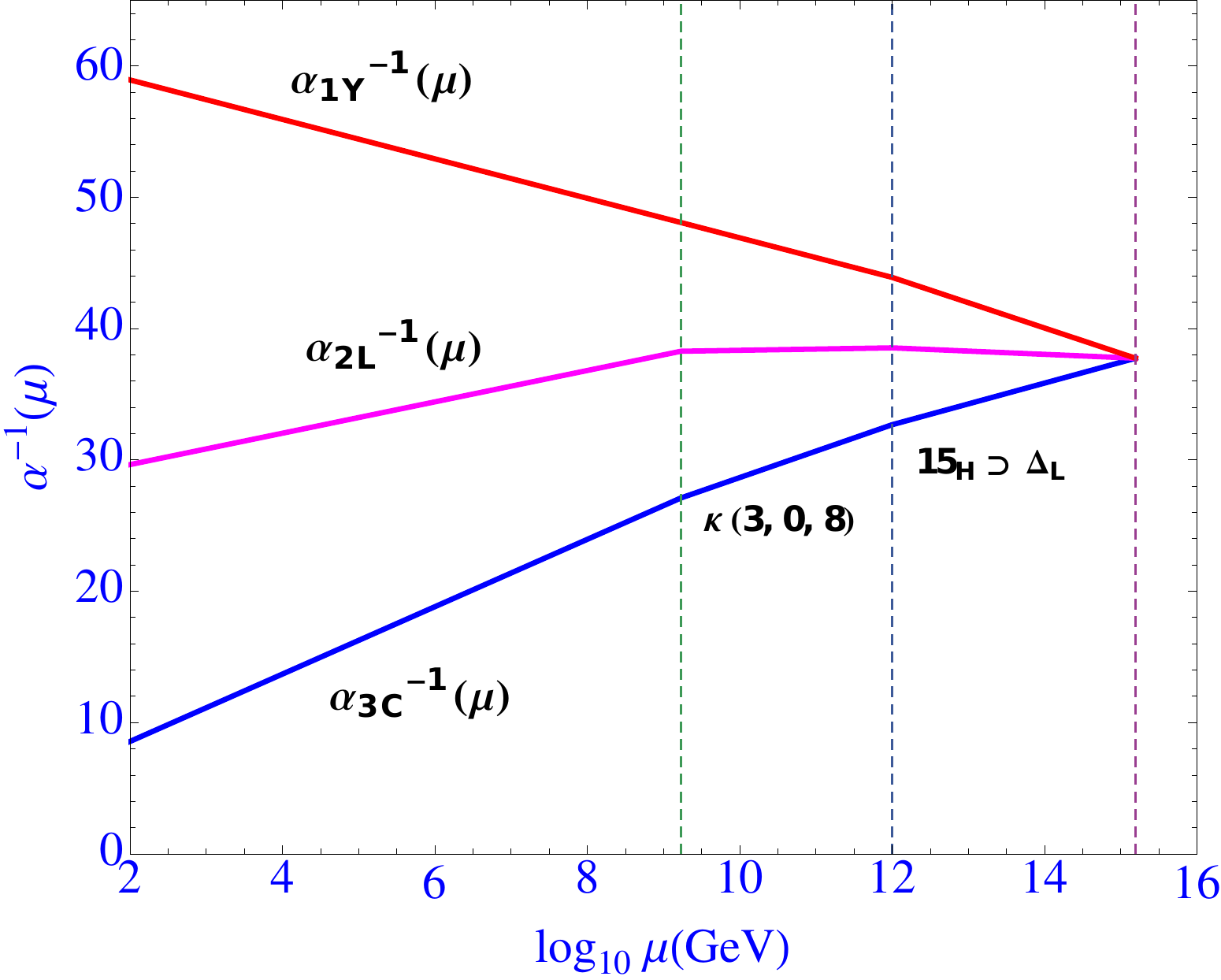} 
\caption{Unification of SM gauge couplings in the presence of $\kappa(3,0,8)$ 
at $M_{\kappa}=10^{9.23}$ GeV and ${15}_H \subset SU(5)$ at $M_{15}=10^{12}$ GeV as  discussed in the text. The vertical dashed lines represent 
the intermediate scale masses  and GUT scales. }
\label{fig:gcu2}
\end{center}
\end{figure}

 In Fig. \ref{fig:gcu2} we have shown evolution of inverse fine
 structure constants of three gauge couplings of SM
 against mass scales depicting precision unification at $M^0_U  =
 10^{15.2}$\, {\rm GeV}.  
\subsubsection{Implications for Lepton Number and Flavor Violations}
It is evident from eq.(\ref{eq:bound}) and  eq.(\ref{eq:massscaleseta}) that
the numerical lower bound on the masses of three members of the
triplet in $\Delta_L(3, -1, 1)$ is \\

$M_{\Delta^0} \simeq M_{\Delta^-} \simeq M_{\Delta^{--}} \ge  10^{9.23} {\rm GeV}$. \\

Out of these we have discussed in Sec.\ref{sec:numass} how the mediation of 
$M_{\Delta^0}$ gives type-II seesaw contribution to neutrino masses
matching with available neutrino oscillation data at
$1\sigma$-$3\sigma$ levels for all types of hierarchies: NH, IH and QD.
As a result the Higgs-Yukwa interaction and induced VEV of the neutral
component of the triplet in the present SU(5) model  gives similar predictions as in the
triplet extended SM based analyses  \cite{Valle:1982} or in the left-right symmetric models and SO(10) with large $W_R$ boson mass \cite{nurev2,RNM-gs:1981,bpn-mkp:2015}.
Currently a
number of experimental
investigations are underway to detect the double beta decay process
that would establish Majorana nature of neutrino. The most important difference from
such SM based phenomenological analyses is that in the present SU(5) model with type-II seesaw   
all the parameters of the neutrino oscillation data are theoretically predicted by
the seesaw mechanism. Even though $M_{\Delta^0}\ge 10^{9.23}$ GeV, it
  predicts the double beta decay lifetime close to the observable
  limit of $\tau_{\beta\beta} \sim 5\times 10^{25}$ yrs. for QD type light neutrino masses
  ${\hat m}_i\sim 0.2$ eV.  On the other hand for NH type of hierarchy
  the predicted decay rate is much lower with  lifetime 
$\tau_{\beta\beta} >> 10^{29}$ yrs.
Another theoretical contribution to the double beta decay process is
due to the mediation of the doubly charged component
$\Delta^{--}$ through the physical process $W^-W^- \to \Delta^{--} \to
e^-e^-$ which is  negligible because of additional damping of the amplitude caused by the inverse square of its heavy
mass $M_{\Delta^{--}} \ge 10^{9.23}$ GeV. 
The charged component $\Delta^-$  also mediates a 
new loop contributions to lepton flavor violating processes such as
$l_{\alpha} \to  l_{\beta}+ \gamma$. Again,
because of heavy triplet mass the respective contribution to branching ratio turns out to
be much smaller than the corresponding prediction with $SM$
(supplemented by the oscillation data): $Br.(\mu\to e\gamma) <
10^{-53}$ \cite{nurev2,bpn-mkp:2015}.  Similarly the tree level mediation of the LFV process $\mu
\to ee{\bar e}$ by $\Delta^{--}$ is severely damped out compared to
the loop mediated $W-$ boson contribution. 
\subsection{Threshold Effects on the GUT Scale}
In the single step breaking model discussed in this work, GUT threshold effects
due to superheavy degrees of freedom in different SU(5) representations  are
expected as major sources of uncertainties on unification scale and  proton
lifetime prediction. We have estimated the threshold uncertainties following the
 partially degenerate assumption  introduced in \cite{rnm-mkp:1993}
 which states that the superheavy components belonging to the same GUT
 representation are degenerate with a single mass scale.

 The analytic formulas for GUT threshold effects on the unification scale, intermediate scale
 and GUT fine structure constant are

\begin{align}
\Delta_{I}^{\kappa} =\Delta\ln{M_{\kappa}\over M_Z}& ={1\over 2244}(123\lambda_{2L}-215\lambda_Y+92\lambda_{3C}), \nonumber \\
\Delta_{I}^{U} =\Delta\ln{M_U \over M_Z}& ={5\over 3366}(3\lambda_{2L}+13\lambda_Y
-16\lambda_{3C}), \nonumber\\
\Delta_{I}^{\alpha_{G}} =\Delta({1 \over \alpha_G})&={1\over 80784\pi}(-948\lambda_{2L}-2425\lambda_Y+ 5056\lambda_{3C}).
\label{eq:theq1} 
\end{align}

In eq.(\ref{eq:theq1}) $\lambda_i,i=2L,Y, 3C$ are matching functions  due to superheavy scalars (S) and gauge bosons (V) to the three gauge couplings,
\be
\alpha^{-1}_i (M_U)=\alpha^{-1}_G-\frac {\lambda_i(M_U)}{12\pi},\label{eq:matching} 
\ee
\ba
\lambda_{i}^S(M_U) & =&\sum_{j}Tr\left(t_{iSj}^2\hat{p}_{Sj}\ln{M_j^S \over M_U}\right),\nonumber \\
\lambda_{i}^V(M_U) & =&\sum_{l} Tr\left(t_{iVl}^2\right)
-21\sum_lTr\left(t_{iVl}^2\ln{M_l^V \over M_U}\right),\label{eq:thform2}
\ea
 where $t_{iS}$  and  $t_{iV}$ represent the matrix representations due to 
broken generators of scalars and gauge bosons. The term $\hat{p}_{Sj}$
denotes the projection operator  that removes the Goldstone components from 
the scalars  contributing to spontaneous symmetry breaking. 

    The decomposition of different SU(5) representations under $G_{213}$ with respect to 
their superheavy components and values of corresponding matching functions are presented in 
Table.\ref{tab:decompsu5}

 \begin{table}[h!]
\caption{Superheavy components of  
SU(5) representations under the SM gauge group $G_{213}$ used to
estimate GUT threshold effects.}
\centering
\vskip 0.5cm
\begin{tabular}{| p{3.8cm} | p{3.5cm} | p{3.5 cm} | }
\hline
$SU(5)$representations & $G_{213}$ submultiplet & ($\lambda_{2L},~ \lambda_{1Y},~\lambda_{3C}$)\\ \hline
 $5_H$  & $ C_1(1,-1/3,3)$  & ($0,~{2 \over 3},~1$)\\ \hline  
$ 24_H $ & $  D_1(3,0,1)$  & ($2,~0,~0 $)\\
        & $ D_2(1,0,8)$  & ($0,~0,~3 $) \\ \hline
 $75_H$ & $ E_1(1,10/3,3)$ & ($0,~5,~{1\over 2} $) \\
       & $E_2(2,5/3,3) $ & (${3 \over 2},~{5 \over 2},~1 $) \\
       & $ E_3(1,-10/3,\bar{3})$ & ($0,~5,~{1\over 2} $)\\
       & $ E_{4}(2,-5/3,\bar{3})$ &(${3 \over 2},~{5 \over 2},~1 $) \\
       & $ E_{5}(2,-5/3,\bar{6})$ &  ($3,~5,~5 $) \\
       & $E_{6}(2,5/3,6) $ &  ($3,~5,~5 $) \\
       & $E_{7}(1,0,8)$ &  ($0,~0,~3 $) \\ \hline 
 $15_H$ & $ \Delta_L(3,-1,1) $ &  ($4,~{18 \over 5},~0 $) \\
       & $H_2(2,1/6,3)$ &   ($3,~{1 \over 5},~2 $) \\
       & $H_3(1,2/3,6)$ &  ($0,~{16 \over 5},~5 $) \\ \hline
$ 24_V $ & $V_1(2,-{5 \over 6},3)$ & (${3 \over 4},~{5 \over 4},~{1 \over 2}$)\\
       & $V_2(2,{5 \over 6},\bar{3})$ & (${3 \over 4},~{5 \over 4},~{1 \over 2}$) \\ \hline 
\end{tabular}
\label{tab:decompsu5}
\end{table}

Using the values of  matching function $\lambda^i(M_U)$ from the Table.{\ref{tab:decompsu5}} in eq.(\ref{eq:theq1}) we estimate corrections to different mass scales due to superheavy masses 
 as 
\begin{align}
\Delta\ln{M_{\kappa} \over M_Z}& = 0.0026738\eta_{5}+0.23262\eta_{24}
-1.24599\eta_{75}, \nonumber \\
\Delta\ln{M_U \over M_Z}& = -0.0160428\eta_{5}-0.0623886\eta_{24}
+1.142602\eta_{75}, \nonumber \\
\Delta({1 \over \alpha_G})& = 0.0160999\eta_{5}+0.0522951\eta_{24}
+0.0462547\eta_{75}. \nonumber \\
\end{align}
 
Maximising the uncertainty in $M_U$ leads to  ~~~~~~~~
\ba
\Delta \ln ({M_U \over M_Z})&=&\pm 0.22103\eta_{SH} ,~~\nonumber\\
\Delta \ln ({M_{\kappa} \over M_Z})&=& \pm 1.48128\eta_{SH} , ~~\nonumber\\
\Delta \left({1 \over \alpha_G}\right)&=&\pm 0.02214\eta_{SH}.\label{eq:maxthsh} 
\ea 
where $\eta_{SH}=\ln ({M_{SH} \over M_U})$ and $M_{SH}/M_U= n(1/n)$ with
 plausible allowed values of real number $n=1-10$ . 

We also note that the degenerate superheavy gauge bosons contribute a
 significant correction to unification scale
\begin{equation}
 \left(\frac{M_U}{M_U^0}\right)_{V}=10^{\pm 0.65508}. \label{muthV}
\end{equation} 

 Adding all corrections together we obtain
\begin{equation}
M_U=10^{15.2312\pm 0.11\pm 0.221\eta_S\pm 0.655\eta_V} {\rm GeV}. \label{MUtotal}
\end{equation}   
The first uncertainty ($\pm 0.11$) represents uncertainty in input parameters given in eq.(\ref{eq:inputpara}).

\section{Proton Lifetime Prediction}\label{sec:plife}
Currently the measured value on the lower limit of the proton  life
time for the decay mode $p\to e^+\pi^0$ is \cite{Shiozawa:2014,JCP:2017,sk-hk1,sk-hk2,Abe:2017}
\begin{eqnarray}
&&\tau_p^{expt.}~\ge ~1.6\times 10^{34}~~{\rm yrs.} \label{taupexpt}        
\end{eqnarray}

 Including strong and electroweak renormalization
effects on the ${\rm d}=6$ operator and taking into account quark mixing, chiral symmetry breaking
effects, and lattice gauge theory estimations, the decay rates
  are \cite{Nath-Perez:2007,Babu-Pati:2010,Bajc}, 
\be  
\Gamma(p\rightarrow e^+\pi^0)
=(\frac{m_p}{64\pi f_{\pi}^2}
\frac{{\alpha_G}^4}{{M_U}^4})|A_L|^2|\bar{\alpha_H}|^2(1+D'+F)^2\times R,
\label{width}
\ee
where $ R=[A_{SR}^2+A_{SL}^2 (1+ |{V_{ud}}|^2)^2]$ for $SU(5)$, $V_{ud}=0.974=$ 
 the  $(1,1)$ element of $V_{CKM}$ for quark mixings, and
$A_{SL}(A_{SR})$ is the short-distance renormalization factor in the
left (right) sectors.  In eq.(\ref{width}) $A_L=1.25=$
long distance renormalization factor but  
$A_{SL}\simeq A_{SR}=2.542$. These are numerically estimated by
evolving the ${\rm dim.} 6$ operator for proton decay by using the
anomalous dimensions of ref.\cite{Buras:1978} and the beta function
coefficients for gauge couplings of this model. In eq.(\ref{width})  
 $M_U=$ degenerate mass of  superheavy gauge bosons, $\bar\alpha_H =$
hadronic matrix elements, $m_p =$proton mass
$=938.3$ MeV, $f_{\pi}=$ pion decay 
constant $=139$ MeV, and the chiral Lagrangian parameters are $D=0.81$ and
$F=0.47$. With $\alpha_H= \bar{\alpha_H}(1+D'+F)=0.012$ GeV$^3$ estimated from 
lattice
gauge theory computations  \cite{Aoki:2007,Munoz:1986}, we obtain  $A_R \simeq A_LA_{SL}\simeq
A_LA_{SR}\simeq 2.726$ and the expression for the
 inverse
decay rate is
\begin{equation}
\Gamma^{-1}(p\rightarrow e^+\pi^0) 
 =
  \frac{4}{\pi}\frac{f_{\pi}^2}{m_p}\frac{M_U^4}{\alpha_G^2}\frac{1}{\alpha_H^2
    A_R^2}\frac{1}{F_q},\label{taup}
\end{equation}
where the GUT-fine structure constant $\alpha_G=0.0263$ and the
factor  $F_q=(1+(1+|V_{ud}|^2)^2)\simeq 4.8$. 
This formula has 
the same form as given in \cite{Babu-Pati:2010} which has been
modified here for the SU(5) case.

 Using  the estimated values of the model parameters, eq.(\ref{taup}) gives,
\begin{eqnarray}
&&\tau_p^{SU(5)}\simeq  10^{33.110\pm 0.440 \pm 0.884|\eta_S|\pm
    2.62|\eta_V|}~~{\rm yrs}.\label{taupnum}
\end{eqnarray}

Numerical estimations on proton lifetime are shown in Table
\ref{tab:taupdeg1} for different splitting factors of superheavy masses. 
\begin{table}[h!]
\caption{ Upper limits on predicted proton lifetime as a function of superheavy scalar (S) and gauge boson(V) mass splittings as defined in the text. The factor
  $10^{\pm 0.44}$ represents  uncertainty due to input parameters.}
\centering
\begin{tabular} {|p{1.5 cm} |p{1.5 cm }|p{2.7 cm}|p{1.5 cm }|p{1.5 cm}|p{2.7 cm}|}  
 \hline
 ${M_S \over M_U} $ & ${M_V \over M_U}$& $\tau_P (yrs)$ & ${M_S \over M_U} $
 & ${M_V \over M_U} $ & $\tau_P (yrs)$   \\ \hline
 $10$ & $1$ & $9.77\times10^{33\pm 0.44}$ & $5$ & $5$ & $3.59\times10^{35\pm 0.44}$ 
\\ \hline
 $10$ & $2$ & $6.00\times10^{34\pm 0.44}$ & $3$ & $6$ & $3.68\times10^{35\pm 0.44}$
 \\ \hline
 $8$ & $3$ & $1.42\times10^{35\pm 0.44}$ & $1$ & $10$ & $5.32\times10^{35\pm 0.44}$
 \\ \hline
 $6$ & $4$ & $2.35\times10^{35\pm 0.44}$ & $20$ & $1$ & $1.80\times10^{34\pm 0.44}$
 \\ \hline
\end{tabular} 
\label{tab:taupdeg1}
\end{table}

It is interesting to note that, despite three Higgs representations ${5}_H,{24}_H, {75}_H$,  major contribution to threshold uncertainty in the model is only due to superheavy gauge bosons.When all superheavy gauge boson masses are identically equal to $M_U$,   superheavy scalar mass splitting by a factor 20(1/20) from the GUT scale gives $\eta_{S}=1.3(-1.3)$ leading to $[\tau_p]_{max}= 1.80\times 10^{34}$ yrs. which is consistent with the current experimental bound.
         
.
\section{Scalar Dark Matter in SU(5)}\label{sec:wimpdm}
\subsection{Phenomenological and Experimental Constraints} 
The existence of dark matter(DM) in our galaxy has
been established beyond any doubt through its gravitational effects by
numerous observations\cite{BH:2016}. Hence the hunt for DM has been
assumed paramount importance for the particle physics community to
understand its nature in particular and that of the universe in
general. To this end, experiments using a wide range of approaches are
being pursued worldwide and giving a large spectrum of interpretations
of the DM candidates with masses ranging from a few eV to PeV or
even beyond, from axions to wimpzillas and decaying dark matter.

Our motivation in this section is to explore whether SU(5) model can
 accommodate a scalar singlet ($=\xi$) as a candidate DM which might be instrumental in contributing to the observed relic density or  may be detected
 through ongoing direct or indirect search experiments. The local DM density is observed with some uncertainty to be $0.4$ $\rm GeV/\rm cm^3$\cite{Catena:2010}.
Earlier measurements by WMAP \cite{spergel:2007} and more recent observation
by PLANCK satellite\cite{Planck15}  indicate $85\%$ of matter
content of the Universe to be DM with its relic density
\begin{equation}
\Omega_{\rm dm}h_{\rm Hubble}^2=0.1198 \pm 0.0026 \label{eq:relic}
\end{equation}
where $h_{\rm Hubble}$ is the Hubble parameter. 
Various attractive models have
been proposed to explain
the observed  relic density of  dark matter  and its
stability with half life greater  than the age of the
universe , $\tau_{DM}> 10^{17}$s. Attempts in this direction
  include addition of scalar or fermionic dark matter
  candidates to the RH neutrino (RH$\nu$) extended SM. Following the  work of 
Lee-Weinberg \cite{Lee-Weinberg:1977} and   in big-bang cosmology, a weakly interacting massive particle (WIMP) has enjoyed a
special status as a DM candidate 
  as  it can naturally explain the observed
 relic density. Model independent upper bound on the WIMP DM  mass
has been also derived from perturbative unitarity \cite{Griest:1990}
with $M_{\rm WIMP}\le
100$ TeV. Recently extensive investigations have been made to explore possible
 special symmetries underlying the dynamics of DM \cite{Queiroz}.    

\subsubsection{Direct Detection of Dark Matter}
Since DM particles are electrically neutral and cosmologically stable,
they are referred as missing energy at colliders where searches for DM
mainly focus on the detection of visible signals like jets and charged
leptons. At colliders we can study DM either through investigating its
direct detection signals or indirect detection signals. The scalar
singlet DM in our model may be discovered through direct and indirect
signals. In particular, XENON1T experiment may discover or rule out
the scalar singlet DM for reasonable values of DM mass and Higgs
portal coupling, rejecting its non-perturbative values higher than 1.5
TeV\cite{Cline:2015}.

Several terrestrial experiments like CDMS\cite{CDMS:1,CDMS:2},DAMA/NAI\cite{DAMA:1,DAMA:2},
XEXNON100 \cite{XENON} and LUX \cite{LUX:2016}
are still going on around the globe for direct detection of dark
matter. These underground detectors  are constructed using various
targets made up of Xe, Ge, NaI etc. in an attempt to explore either
electronic or nuclear scatterings at low energies. In this case, the
recoil energy is usually observed from the scattering between DM
particles and nucleons\cite{Goodman:1985} or from scattering between
electrons and dark matter. The direct search experiments, XENON100 \cite{XENON} and
LUX, predict an upper bound in the $M_{DM}-\sigma_{DM}$ plane where
$\sigma_{DM}$ represents DM elastic scattering cross-section and
$M_{DM}$ stands for DM mass. These experiments furnish very stringent
bounds on dark matter-nucleon scattering cross-section for different
DM masses. For example, LUX and XENON100 experiments  predict similar
DM-nucleon cross section bound at around $10^{-44} \rm cm^2$ for a DM
mass of $1000$ GeV whereas XENON1T search predicts a smaller cross
section bound $2\times10^{-46} \rm cm^2$ for the same DM mass keeping the
DM relic density in the right ballpark \cite{Laura:2016,Schumann:2015} . A concise review of current status of scalar singlet dark matter is
available in \cite{Garg:2017} where references to most of the recent
 experimental and  phenomenological investigations are available.
In general, for elastic scattering of a DM  particle off nucleons, either a
standard Higgs or a $Z$-boson exchange is needed in the t-channel of the dominant tree diagrams. Even though the singlet scalar DM $\xi(1,0,1)$ has no gauge interaction, still it can elastically scatter off nucleons in direct search experiments through Higgs exchange via quartic Higgs portal interaction  
\begin{equation}
V_{\rm Port}=\frac{\lambda_{\phi \xi}}{2}
\phi^{\dagger}\phi\xi^2 + h.c. \label{eq:portal}
\end{equation} 
where the standard Higgs VEV and the portal quartic coupling $\lambda_{\phi \xi}$
contribute directly to the cross section in the lowest order.

Although till today no signals in direct detection experiments have
been observed except for the controversial DAMA modulation signal, direct detection searches still have the potential to unravel the mystery of DM because of the fact that if a signal is observed, we can correlate the scattering cross section and mass of the DM particle with its local density.

\subsubsection{Indirect Detection of Dark Matter}
 In  indirect dark matter detection(IDMD) experiments, the DM
 particles may annihilate or decay to standard model particles or
 other exotic final states in a region of high DM density and finally
 manifest as a visible signal in form of gamma rays, cosmic rays,
 neutrinos and positrons or anti-particles. Such events are expected
 to exhibit excesses over the desired abundance of the particles in
 the cosmos. The IDMD searches like Fermi-LAT\cite{Fermi-LAT}, AMS\cite{AMS}, HESS\cite{HESS}, MAGIC\cite{MAGIC}, ATIC\cite{ATIC},
 DAMPE\cite{DAMPE}, PLANCK\cite{Planck15}, ICECUBE\cite{IceCube:2013,IceCube:2014 } etc   basically look for
 these excesses in the
 universe to confirm the detection of DM annihilation. For example, DM
 could be detected through the observation of neutrino fluxes  by
 ICECUBE telescope arising from annihilation dark matter. The IceCube
 neutrino events have been recently interpreted to be consistent with
 decaying dark matter mass in the PeV range or larger.

Recently IDMD searches gave several hints for DM detection like lines
at 3.5 KeV \cite{Bulbul:2014,Boyarsky:2014}, 130
GeV\cite{Bringmann:2012,Weinger:2012} and the gamma ray excess from
the galactic centre\cite{Daylon:2014}. However, no conclusive and
consistent information has emerged so far. These signals have been
attributed to either astrophysical sources or instrumental effects\cite{Jeltema:2014,Petrovic:2014}. 

 Recent data from LUX-2016
and Fermi-LAT \cite{LUX:2016,Fermi-LAT} have constrained the DM mass
as well as its unknown Higgs portal coupling. It can be shown that $\lambda_{\phi \xi}
\sim {\cal O} (0.01)$ to generate the right relic density with low mass
$\xi$ of order 50 GeV. On the other hand direct DM searches from the
 LUX-16 data has ruled out the
existence of scalar DM $\xi$ over a wider mass range $M_{\xi}\simeq 70-500$ GeV.
  In summary, the scalar dark matter mass can be on the lower side
\begin{equation}
 M_{\xi} < 60 {\rm GeV}, \label{eq:ub} 
\end{equation}
contributing prominently to relic density,
 or on the higher side
\begin{equation}
100 {\rm TeV}\,\, >  M_{\xi} > 500\,\, {\rm GeV}. \label{eq:lb}
\end{equation}
In eq.(\ref{eq:lb}) the LHS is due to the perturbative unitarity bound
\cite{Lee-Weinberg:1977} and the RHS due to \cite{LUX:2016}.


\subsection{Embedding in SU(5)}
Besides the SU(5) Higgs representations ${5}_H,{24}_H,{15}_H$ and ${75}_H$,
we further extend its scalar sector by the scalar singlet DM
$\xi(1,0,1)$ which we assume to be also a SU(5) singlet. Obviously it
has no direct gauge boson interaction of any kind. But it has
interaction with SM Higgs through Higgs portal of the type shown in
eq.(\ref{eq:portal}). Then it can have gauge interaction in higher orders.     
In any theoretical  model, the stability of DM must be ensured such
that its lifetime is longer than the lifetime of the universe. Usually
a discrete symmetry $Z_2$ is imposed to safeguard the stability.

We assign all the fermions in ${\bar 5}_F$, ${10}_F$, and consequently
the SM fermions, to possess $Z_2=-1$. The Higgs representations
${5}_H,{24}_H,{15}_H$ and ${75}_H$ are assigned $Z_2=+1$. Needless to
mention the SM Higgs doublet $\phi$, $\kappa(3,0,8)$, and
$\Delta_L(3,0,1)$ have the same value of  $Z_2=+1$. Out of all the
scalars only the DM singlet scalar is assigned odd value of
$Z_2=-1$. This assignment prevents direct Yukawa interaction of 
$\xi$ and ensures its desired stability.
\section{Vacuum Stability in SU(5) Through Scalar DM}\label{sec:vacstab} 

Despite the above predictions  on neutrino masses and mixings,
coupling unification, and proton lifetime, the SU(5) model with Higgs
representations still has
the vacuum instability problem. This problem in the SM arises as the
standard Higgs potential solely controlled by the standard Higgs field
becomes unstable for large values of the field at scales $\mu \ge
M_{\rm Inst.} =
5\times 10^{9}$ GeV. As there is no other field so far in the extended
SU(5) model for $\mu < M_{\Delta}(=10^{12}-10^{15} {\rm GeV})$ to
couple through its Higss portal, the instability problem turns out to
be similar to SM. As we have embedded the scalar singlet DM candidate
in SU(5) we now investigate the possibility of resolving the
vacuum instability through Higgs partial interaction \cite{Elias-Miro:2012,Lebedev:2013,Garg:2017}. 

\subsection{RG Equations and Parameters for Higgs Potential} 
 As noted above the standard model Higgs potential
\begin{equation}
  V_{SM}=-{\mu_{\phi}}^2 \phi^{\dagger}\phi+ \lambda_{\phi}(\phi^{\dagger}\phi)^2
\end{equation}
develops instability as the Higgs quartic coupling $\lambda_{\phi}$ runs 
negative at an energy scale $10^9-10^{10}$GeV by the renormalization group 
running. 
Apart from other interesting suggestions \cite{Elias-Miro:2012,
  Lebedev:2013} an alternative popular solution  to the vacuum instability problem is to 
extend the SM by a gauge singlet real scalar($\xi$) which gives positive contribution to the Higgs quartic coupling and prevents it from becoming negative \cite{Elias-Miro:2012,gond:2010,chen:2012,khan:2014}.  
It is worth mentioning that this scalar singlet  can act as potential dark 
matter candidate termed as weakly interacting massive particle(WIMP) with an 
extra discrete symmetry $Z_2:\xi \rightarrow -\xi$ imposed on it. The scalar singlet
 is odd under $Z_2$ symmetry while all other scalars are even and SM
 fermions are  odd under this
 symmetry. Hence it can not couple to SM particle and become
 stable. This also matches the discrete symmetry properties of SU(5)
 representations discussed above. Thus it can serve as a suitable WIMP
 dark matter particle which is also identified as the SU(5) singlet
 scalar. The unbroken discrete symmetry of the
 singlet scalar upto the Planck scale has two important
 consequences:(i) The $\xi$ VEV is forbidden,(ii) The modified SM
 potential develops VEV and minima only due to the SM Higgs. 
 The  scalar $\kappa (3,0,8)$ has no coupling with $\phi$. Even if
 $\Delta_L$ and some of its associates have coupling with $\phi$,
 because of their heavy mass, $M_{\Delta} >> M_{W}$, they are treated
 to have decoupled from the Lagrangian at energy scales below $\mu
 \sim M_{\Delta}$.\\    

\par\noindent{\large\bf{\underline{ $\mu < M_{\Delta}$}}}\\ 

The potential of the model is modified in presence of the scalar singlet and a new term arises  due to interaction of SM doublet ($\phi$) with scalar singlet ($\xi$)
and self interaction of $\xi$ 
\begin{equation}
  V(\xi,\phi)= V_{\rm SM}+{\lambda_{\phi \xi} \over 2}\phi^{\dagger}\phi \xi^2 + {\mu_{ \xi}^2 \over 2}\xi^2+
        {\lambda_{ \xi} \over 24}\xi^4 
\label{pots}
\end{equation}
 where $\lambda_{\xi}$ is dark matter self-coupling ,~$\lambda_{\phi \xi}$  is standard Higgs and  extra Higgs scalar interaction coupling or Higgs portal coupling
 and $\mu_\xi$ is quadratic coupling of extra Higgs scalar. From electroweak scale, upto $\mu=10^{12}$ GeV the effective potential is $ V^\prime(\xi,\phi)=V_{SM}+ V(\xi,\phi)$.\\   

\par\noindent{\large\bf{\underline{ $\mu > M_{\Delta}$}}}\\ 

The introduction of the scalar triplet $\Delta_L$ of mass
$M_{\Delta}\sim 10^{12}$ GeV, changes the Higgs potential further
  by additional terms $V(\phi,\Delta_L)$ 
(arising out of interaction of SM doublet with scalar triplet and self
  interaction of scalar triplet) and $V(\xi,\Delta_L)$ (arising out of interaction of scalar singlet DM and scalar triplet)
\begin{equation}
V(\xi,\phi,\Delta_L)= V(\xi,\phi)+V(\phi,\Delta_L)+V(\xi,\Delta_L)
\end{equation}
where
\begin{eqnarray}
V(\phi,\Delta_L)& = & M_\Delta^2 {\rm Tr}(\Delta_L^\dagger \Delta_L)
          + \frac{\lambda_1}{2} \left[ {\rm Tr}(\Delta_L^\dagger \Delta_L) \right]^2+ 
           \frac{\lambda_2}{2} \left( \left[ {\rm Tr}(\Delta_L^\dagger \Delta_L) \right]^2	
           - {\rm Tr} \left[ (\Delta_L^\dagger \Delta_L)^2 \right] \right) \nonumber \\
         &&+ \lambda_4 (\phi^\dagger \phi) {\rm Tr}(\Delta_L^\dagger \Delta_L)
         + \lambda_5 \phi^\dagger [\Delta_L^\dagger, \Delta_L] \phi 
          +\left( \mu_\Delta\tilde{\phi}^\dagger(\frac{\vec{\tau}.\vec{\Delta_L}}{\sqrt{2}})^\dagger
\phi +{\rm h.c}. \right)\nonumber\\
V(\xi, \Delta_L) &=& \lambda_{\xi \Delta_L} (\xi^\dagger \xi) (\Delta_L^\dagger \Delta_L)~.
\end{eqnarray}


Sufficiently below the mass scale $\mu=M_{\Delta}=10^{12}$ GeV, our model has two scalars:the first one is the SM Higgs($\phi$) given by
  $\phi={1\over\sqrt{2}}\left(\phi^+,~ v+h+i\phi^0 \right)^T $  
 and the second one is extra scalar singlet($\xi$) added to the SM. The mass of the extra singlet is given by \\
\begin{equation}
  M_{DM}^2=\mu_{\xi}^2 +{\lambda_{\phi \xi}\over 2} v^2.\label{eq:DMmass}  
\end{equation}
We use the standard Higgs mass  $m_{h}=125$GeV.

 Direct detection experiments \cite{LUX:2016,Fermi-LAT}
 impose constraints on the Higgs portal coupling
$(\lambda_{\phi \xi})$ and dark matter mass\cite{khan:2014,pa:2017}
 derived from observed DM relic density

\be
M_{DM}\sim 3300\times\lambda_{\phi \xi}
\ee
or
\be
\lambda_{\phi \xi} \sim 0.0003\times M_{DM}. \label{eq:con} 
\ee
for $M_{DM}>> m_{\rm top}$. To be consistent with eq.(\ref{eq:lb}) we
 use $M_{DM} =m_{\xi}\sim 1$TeV through out this work. Similar
 analysis can be carried out for all values of  DM mass $> 500 $ GeV. 

These constraint on  $\lambda_{\phi \xi}$ given in eq.(\ref{eq:con})
can be also considerably relaxed
if there are more than one WIMP DM candidate of the same or different
species including fermions \cite{Strummia:2006,pnsa:2016}. 

\subsection {RG Evolution of Quartic Coupling}
 Like other couplings of every non-Abelian gauge theory, it is well known that the SM Higgs potential is modified by 
 quantum corrections determined by perturbative renormalization group
  equations (RGEs) for its running couplings $\eta(\mu)$              
\be
{d\eta \over dt}=\sum_{j}{\eta^{(j)} \over (16\pi^2)^j}
\ee
where $t=\log\mu$,\, $\mu$ is renormalization scale,~ $\eta (\mu)=$different 
couplings (quartic or gauge, or others) at scale $\mu$ and $j=j^{th}$~loop order.
 The one-loop RG-coefficients of different couplings are presented in the 
Appendix. For the stability of the Higgs potential (eq.\ref{pots}),
the value of self coupling including corrections must remain
positive throughout the course of its evolution up to the Planck scale. 

\begin{center}
\begin{figure}[h!]
\includegraphics[scale=0.55]{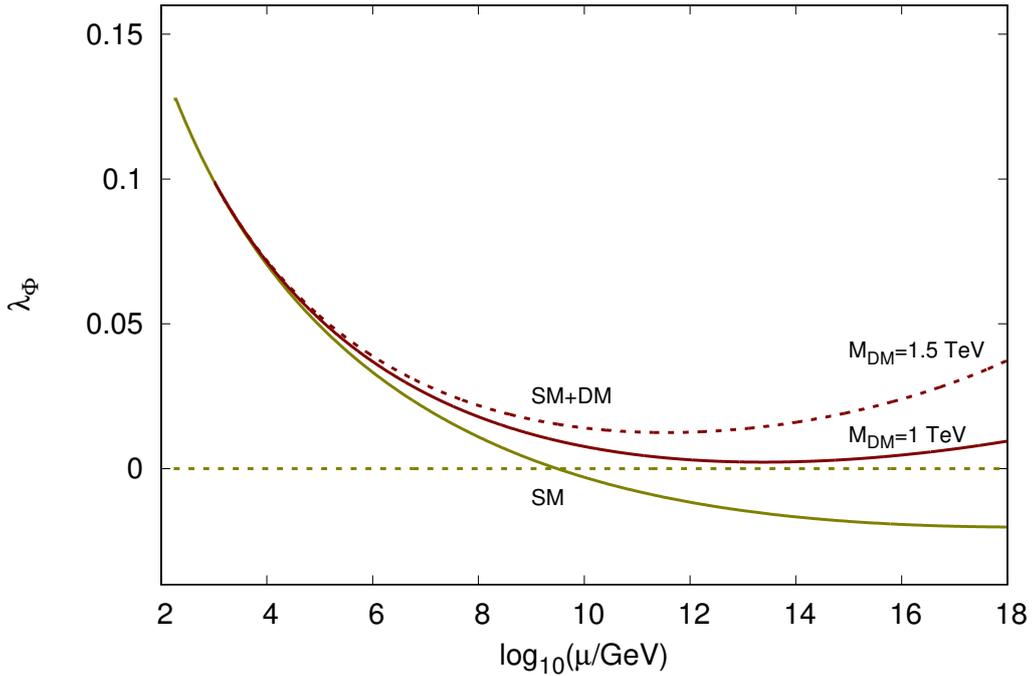}
\caption{Running of Higgs quartic coupling. }
\label{fig:stabm}
\end{figure}
\end{center}

The running of Higgs quartic coupling $\lambda_{\phi}(\mu)$ with
  energy scale $\mu$ is shown in Fig.\ref{fig:stabm}
 where, at first, we have neglected possible threshold effects due to Higgs
 triplet at $\mu=M_{\Delta}$ determined as one of the solutions to
 neutrino oscillation data. Negligible $\Delta_L-$ threshold effect can also
 result for $\mu_{\Delta}<< M_{\Delta}$. 
We have used the initial values of different coupling constants at top
quark mass scale($\mu=m_{\rm top}$) as given in the Table.{\ref{tab:cc}} and subsequently evolved them from $m_t$ to Planck scale with the help of RGEs.

\begin{table}[h!]
\caption{Initial values of coupling constants at top quark mass.}
\centering
\begin{tabular} {|p{1.7 cm} |p{1.4 cm }|p{1.4 cm} |p{1.4 cm }|p{1.4 cm}|p{1.4 cm }|p{1.4 cm}|p{1.4 cm}|}  
 \hline
  Coupling constants & $\lambda_{\phi}(m_t)$& $\lambda_{\xi}(m_t) $ & $\lambda_{\phi \xi}(m_t) $ & $g_{1Y}(m_t)$ & $g_{2L}(m_t)$ &  $g_{3C}(m_t)$ &  $y_{t}(m_t)$   \\ \hline
  Initial values & $ 0.1296 $ & $0.1$ & $0.36 $ & $0.35 $ & $0.64 $& $1.16$ &
$0.94 $   \\ \hline
\end{tabular} 
\label{tab:cc}
\end{table}

From Fig.{\ref{fig:stabm}}, it is clear that the desired quartic coupling remains 
stable upto the Planck scale for $\lambda_{\phi \xi}=0.36$ and
$M_{DM}=1$ TeV . 
\subsection{Higgs Triplet Threshold Effect}
Threshold effect due to heavier Higgs masses which couple to $\phi$
through their portals has been discussed in general 
\cite{Elias-Miro:2012,Lebedev:2013} and in specific cases \cite{Arhib,Haba:2016}. In our
case the Higgs triplet mass used to fit the neutrino oscillation data
is $M_{\Delta}\sim 10^{12}$ GeV and its induced VEV is ${\cal O}(1-10)$ eV. In such a
case the threshold effect caused by the triplet VEV correction term
is \cite{Elias-Miro:2012} 
\begin{equation} 
\Delta \lambda_{\phi}=\lambda_{\phi \Delta}\frac{v_L^2}{M_{\Delta}^2}
 \sim 10^{-36} \label{eq:vevthrs}
\end{equation}
The remaining threshold effect could be due the self energy correction
or the trilinear term $\mu_{\Delta}\Delta_L\phi\phi +h.c$ in the Higgs
potential giving rise to threshold correction to quartic coupling 
\begin{equation}
\Delta \lambda_{\phi} \equiv \lambda_{\rm TH}=\frac{\mu_{\Delta}^2}{M_{\Delta}^2} \label{eq:selfthrs}
\end{equation}

Denoting the effective Higgs quartic coupling by $\lambda^{\prime}(\mu)$
for $\mu\ge M_{\Delta}$ this is related to the quartic coupling
$\lambda_{\phi}(\mu)$ at $\mu=M_{\Delta}$ \cite{Elias-Miro:2012}
\begin{equation}
\lambda_{\phi}(M_{\Delta})= \lambda^{\prime}(M_{\Delta})-\lambda_{\rm
  TH}.\label{
eq:theff}
\end{equation}
This correction comes into play when the running mass scale $\mu \sim
M_{\Delta}$ and larger. 

We point out that the same values of Majorana Yukawa coupling elements
of $Y$
derived in Sec.\ref{sec:numass} are valid  upto a scale factor for a wide range of values of
trilinear coupling mass parameter $\mu_{\Delta} < M^0_{\Delta}$ for
which this threshold effect is well within the perturbative
regime. We note from eq.\ref{eq:vl} that the mass formula gives the
scaling relation
\begin{equation}
Y= Y^0\frac{v^0_L}{v_L}=Y^0\frac{\mu^0_{\Delta}}{\mu_{\Delta}}.\label{eq:scaling}
\end{equation}
where we have used the zero superscript for values at
$\mu_{\Delta}^0=M_{\Delta}=10^{12}$ GeV. Thus, for the values of
neutrino mass and mixing given by the oscillation data, a new set of
elements of $Y$ are derived for any $\mu_{\Delta} < \mu_{\Delta}^0$ by
multiplying all the vales given Table \ref{Y_no} and Table \ref{Y_io} by the same
scale factor $\frac{\mu^0_{\Delta}}{\mu_{\Delta}}$.

\begin{figure}[h!]
\begin{center}
\includegraphics[scale=0.55]{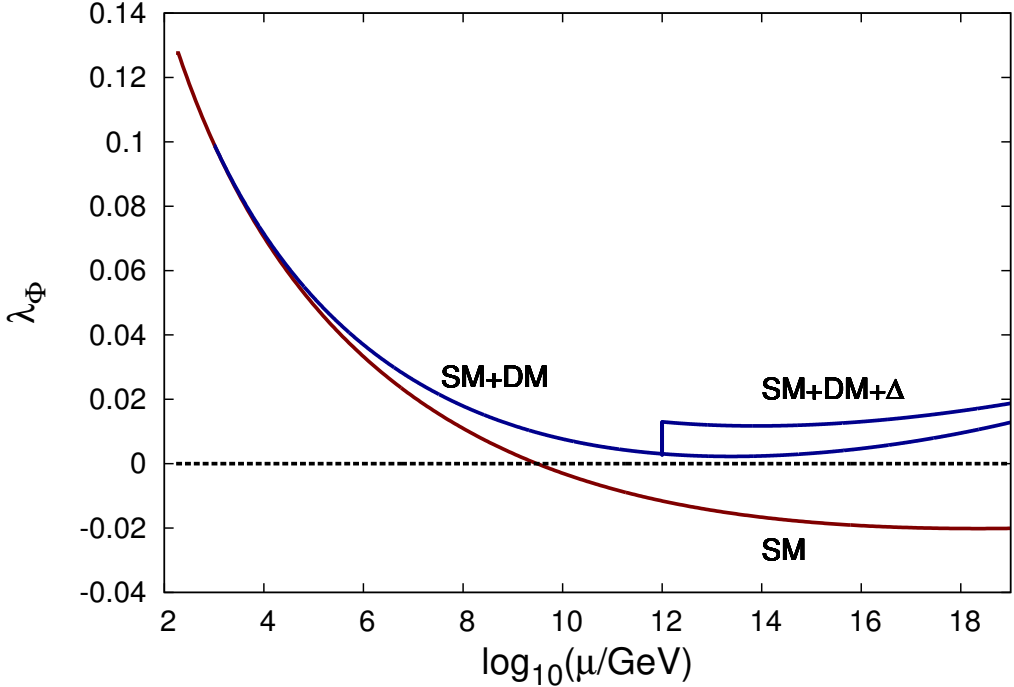}
\caption{Running of standard Higgs quartic coupling including heavy
  triplet scalar threshold effect at $\mu=M_{\Delta }=10^{12}$ GeV derived
  from fits to neutrino oscillation data. The curves labeled as SM,
  SM$+$DM, and SM$+$DM$+$$\Delta$ denote contributions due to SM alone, SM
  plus DM, and SM plus DM plus Higgs triplet threshold effect
  , respectively, as described in the text. The scalar DM mass has
  been fixed at $M_{\rm DM}=1.0$ TeV consistent with LUX:2016
experimental data.} 
\label{fig:vacsdel}
\end{center}
\end{figure}

 In Fig. \ref{fig:vacsdel} we have presented evolution of Higgs quartic
 couplings below and above $\mu=M_{\Delta}$ for $\lambda_{\rm
   TH}=0.1$. Using the notations of Appendix B, we have used the initial values of different coupling constants at scalar triplet mass scale($M_{\Delta}$) as $\lambda_1=\lambda_2=\lambda_4=0.1$ and $\lambda_5=0.1$.

For all the three curves given in Fig. \ref{fig:vacsdel} the scalar DM
mass has been fixed at $M_{\rm DM}=1$ TeV consistent with LUX:2016 data. The curve labeled as
SM$+$DM$+$$\Delta$ includes threshold effect $\lambda_{\rm TH}=0.1$ at
$\mu=M_{\Delta}=10^{12}$ GeV corresponding to $\mu_{\Delta}\sim (1/3) M_{\Delta}$. 
We have checked that even after including the heavy scalar threshold
effect the quartic coupling remains
 perturbatively positive upto the Planck scale for
 $\frac{\mu_{\Delta}}{M_{\Delta}} \simeq 0.5$ \footnote{Denoting
   $\Phi_H={24}_H$,  above the mass
   scale $\mu > M_{\rm GUT}$ we impose the well known discrete symmetry $\Phi_H \to
   -\Phi_{H}$ which is usually assumed in the minimal SU(5) model. Without loss
   of generality we further assume the Higgs portal coupling
   $\lambda_{5_H,{75}_H}$  to be negligible.}.   
   
Thus, the issue of
vacuum stability of SM Higgs potential is resolved through the
embedding of $\xi$ as a WIMP dark matter candidate in SU(5) even after
including the heavy Higgs triplet threshold effect which could be
verified by direct search experiments and LHC.

\section{Summary and Conclusion}\label{sec:sum}
In this work we have attempted to resolve four limitations of the
minimal SU(5) model by extending its scalar sector beyond ${5}_H$ and
${24}_H$. Added presence of ${15}_H$ and ${75}_H$ is noted to account
for precision coupling unification with experimentally verifiable
proton lifetime for $p\to e^+\pi^0$, and type-II seesaw ansatz for
neutrino masses. The left-handed triplet Higgs mass in this model is
bounded from below $(M_{\Delta}=M_{{15}_H}) \ge
(M_{\kappa}=10^{9.23})$ GeV. Proton lifetime is predicted by taking
into account sources of theoretical uncertainties due to GUT threshold
effects and those due to electroweak precision parameters. Type-II
seesaw scale effect on proton lifetime prediction is also
discussed. The limitation due to vacuum stability of the Higgs
potential in SU(5) is resolved by the inclusion of a scalar singlet
near the TeV scale that acts as a WIMP dark matter candidate. All the
fermions and this scalar are assigned to be odd under a dark matter
stabilising $Z_2$ discrete symmetry whereas the SM Higgs is even. The
scalar dark matter mass is consistent with current experimental
LUX-2016 bound on direct search experiments. Renormalization group evolution of SM Higgs quartic coupling modified by Higgs portal coupling of this scalar DM completely alleviates the vacuum instability problem.
We emphasize that no nonstandard Higgs field, except the scalar DM singlet, is present  in this model below the $\kappa$ mass $M_{\kappa}=10^{9.23}$ GeV.
\\
 We thus conclude that  such SM limitations as neutrino mass, coupling
 unification, proton lifetime, WIMP dark matter, and vacuum stability
 can be effectively resolved by extending the scalar sector of SU(5)
 to include ${5}_H,{24}_H,{75}_H,{15}_H$ and ${1}_H$. At present we
 need no extension on the established fermion structure of the SM and
 SU(5) or their minimal gauge structure. The remaining limitations on
 baryon asymmetry generation and/or the possibility of decaying dark
 matter projected to manifest as PeV energy IceCube neutrinos will be
 addressed elsewhere \cite{cmp-BAU,Mainak:2018}.   

\section{APPENDIX: Renormalization Group Equations for
  Higgs Scalar Couplings}\label{sec:app}
The RGEs for scalar quartic couplings\cite{Bajc-gs:2007,Haba:2016} in our model at one loop level are given 
by
\be
16\pi^2{dC \over dt}=\beta_{C}~~(C=\lambda_{\phi},\lambda_{\phi \xi},\lambda_\xi,\lambda_1,\lambda_2,\lambda_4,\lambda_5)
\ee
 where 
\begin{align}
\beta_{\lambda_{\phi}}= & 24\lambda_{\phi}^2+12\lambda_{\phi}y_t^2-6y_t^4-3\lambda_{\phi}(g_{1Y}^2+3g_{2L}^2)+{3\over 8}[2g_{2L}^4+(g_{1Y}^2+g_{2L}^2)^2]+
{\lambda_{\phi \xi}^2\over 2} \\ \nonumber
\beta_{\lambda_{\phi \xi}}=& \{4\lambda_{\phi \xi}+12\lambda_{\phi}+6y_t^2-{3 \over 2}
(g_{1Y}^2+3g_{2L}^2)+\lambda_\xi \} \lambda_{\phi \xi}\\ \nonumber
\beta_{\lambda_{\xi}}= & 3\lambda_\xi^2+12\lambda_{\phi \xi}^2
\label{eq:crge}
\end{align}
  For Standard model RG running in the energy scale $\mu < M_{DM}$, the term
${\lambda_{\phi \xi}\over 2}$ in $\beta_{\lambda_{\phi}}$ in eq.(\ref{eq:crge}) is to be ignored . The RGEs for SM gauge couplings and top quark yukawa coupling at two loop level are given by

\begin{align}
{dy_t \over dt}= & {1 \over 16\pi^2}\left({9 \over 2}y_t^2-{17 \over 12}g_{1Y}^2
 -{9 \over 4}g_{2L}^2-8g_{3C}^2 \right)y_t \\ \nonumber
  +  & {1 \over (16\pi^2)^2} [-{23 \over 4}g_{2L}^4-{3 \over 4}g_{2L}^2g_{1Y}^2+{1187 \over 216}g_{1Y}^4 + 9g_{2L}^2g_{3C}^2+{19 \over 9}g_{3C}^2g_{1Y}^2-108g_{3C}^4 
\\ \nonumber
+& \left({225 \over 16}g_{2L}^2+{131 \over 16}g_{1Y}^2+36g_{3C}^2 \right)y_t^2+6(-2y_{t}^4-2y_{t}^2\lambda_{\phi}+\lambda_{\phi}^2) ] \\ \nonumber
 {dg_{1Y} \over dt}= & {1 \over 16\pi^2}\left({41 \over 6}g_{1Y}^3\right)+{1 \over (16\pi^2)^2}\left({199 \over 18}g_{1Y}^2+{9 \over 2}g_{2L}^2+{44 \over 3}g_{3C}^2-
{17 \over 6}y_t^2\right)g_{1Y}^3 \\ \nonumber
{dg_{2L} \over dt}= & {1 \over 16\pi^2}\left(-{19\over 6}g_{2L}^3\right)+{1 \over (16\pi^2)^2}\left({3 \over 2}g_{1Y}^2+{35 \over 6}g_{2L}^2+12g_{3C}^2-
{3 \over 2}y_t^2\right)g_{2L}^3 \\ \nonumber
{dg_{3C} \over dt}= & {1 \over 16\pi^2}\left(-7g_{3C}^3\right)+{1 \over (16\pi^2)^2}\left({11 \over 6}g_{1Y}^2+{9 \over 2}g_{2L}^2-26g_{3C}^2-
2y_t^2\right)g_{3C}^3
\end{align}
After $\mu=10^{12}$ GeV the scalar triplet $\Delta_L$ is introduced and we use the modified RG equations of $\lambda_\phi$ and other couplings relevant for this 
scalar triplet.
\begin{eqnarray}
 \beta_{\lambda_\phi} &=&
	\lambda_\phi \left[ 12\lambda_\phi - \left( \frac{9}{5}g_{1Y}^2 + 9g_{2L}^2 \right) + 12y_t^2 \right]
	+ \frac{9}{4} \left( \frac{3}{25}g_{1Y}^4 + \frac{2}{5}g_{1Y}^2g_{2L}^2 + g_{2L}^4 \right) \nonumber \\
	&& + 6\lambda_4^2 + 4\lambda_5^2 - 12y_t^4,
\label{beta} \\
 \beta_{\lambda_1} &=&
	\lambda_1 \left[ 14\lambda_1 + 4 \lambda_2
	- \left( \frac{36}{5} g_{1Y}^2 + 24g_{2L}^2 \right)
	+ 4 {\rm tr} \left[T \right] \right]
	+ \frac{108}{25}g_{1Y}^4 + \frac{72}{5}g_{1Y}^2g_{2L}^2 + 18g_{2L}^4 \nonumber\\
	&& + 2 \lambda_2^2 + 4 \lambda_4^2 + 4\lambda_5^2
	- 8 {\rm tr} \left[T^2 \right], \\
 \beta_{\lambda_2} &=&
	\lambda_2 \left[ 12 \lambda_1 + 3 \lambda_2 
	- \left( \frac{36}{5}g_{1Y}^2 + 24g_{2L}^2 \right)
	+ 4 {\rm tr} \left[T  \right] \right]
	- \frac{144}{5}g_{1Y}^2g_{2L}^2 + 12g_{2L}^4 \nonumber\\
	&& - 8 \lambda_5^2 + 8 {\rm tr} \left[T^2 \right],  \\
 \beta_{\lambda_4} &=&
	\lambda_4 \left[ 6 \lambda_\phi + 8 \lambda_1 + 2 \lambda_2 + 4\lambda_4
	- \left( \frac{9}{2}g_{1Y}^2 + \frac{33}{2}g_{2L}^2 \right) 
	+ 6 y_t^2 + 2 {\rm tr} \left[T \right] \right] \nonumber\\
	&& + \frac{27}{25}g_{1Y}^4 + 6g_{2L}^4
	+ 8 \lambda_5^2 - 4 {\rm tr}\left[ T^2 \right], \\
 \beta_{\lambda_5} &=&
	\lambda_5 \left[ 2 \lambda + 2\lambda_1 - 2\lambda_2 + 8 \lambda_4
	- \left( \frac{9}{2}g_{1Y}^2 + \frac{33}{2}g_{2L}^2 \right)
	+ 6 y_t^2 + 2 {\rm tr}\left[T \right] \right]
	- \frac{18}{5}g_{1Y}^2g_{2L}^2 \nonumber\\
	&& + 4 {\rm tr}\left[T^2 \right],
\end{eqnarray}
 where $T$ is defined as $T=Y^\dagger Y $
 and its beta function is expressed through the relation
\begin{eqnarray}
\beta_T =
	T \left[ 6\, T - 3 \left( \frac{3}{5} g_{1Y}^2 + 3 g_{2L}^2 \right)
	+ 2 {\rm tr}[T] \right] .
\end{eqnarray}

\section{ACKNOWLEDGMENT}
M. K. P. acknowledges financial support under the project
SB/S2/HEP-011/2013 from the Department of Science and Technology,
Government of India.  
For financial support from Siksha 'O' Anusandhan (SOA),
Deemed to be University, M. C. acknowledges  a 
Post-Doctoral fellowship and B.S. thanks for a Ph. D. research fellowship.

\end{document}